\documentstyle[12pt,aaspp4,epsfig]{article} 

\newcommand{\vv}[1]{{\bf #1}}
\newcommand{\df}{\delta}

\newcommand{\iras}{{\sl IRAS\/}}
\newcommand{\avg}[1]{{\langle{#1}\rangle}}
\newcommand{\Avg}[1]{{\left\langle{#1}\right\rangle}}

\newcounter{thefigs}
\newcommand{\fignum}{\arabic{thefigs}}

\newcounter{thetabs}
\newcommand{\tabnum}{\arabic{thetabs}}

\begin{document} 

%
%
\title{The Physical Origin of Scale Dependent Bias \\
in Cosmological Simulations}
\vspace{20pt}
\author{Michael Blanton, Renyue Cen, \\
Jeremiah P.~Ostriker, and Michael A.~Strauss\footnote{Alfred P.~Sloan
Foundation Fellow}$^{,}$\footnote{Cottrell Scholar of Research Corporation}}
\vspace{40pt}
\affil{Princeton University Observatory, Princeton, NJ 08544 }
\affil{ blanton, cen, jpo, strauss@astro.princeton.edu}
\vspace{30pt}

%
%

\begin{abstract} 

Using a large-scale hydrodynamic simulation with heuristic criteria for
galaxy formation, we investigate how the galaxy field is related to
physical parameters, such as the mass density and the gas temperature.
In our flat Cold Dark Matter model with $\Omega_0 = 0.37$, we find
that the relation between the galaxy and mass density fields 
is a function of scale. The bias $b(R)\equiv\sigma_g(R)/\sigma(R)$,
where $\sigma_g(R)$ is the variance of galaxy counts in spheres of
radius $R$ and $\sigma(R)$ is the same for mass, varies from 2.6 at 1
$h^{-1}$ Mpc to 1.2 at 30 $h^{-1}$ Mpc. Including the dependence of
the galaxy density on local gas temperature as well as on local mass
density can fully account for this scale dependence. Galaxy density
depends on temperature because gas which is too hot cannot cool to
form galaxies; this causes scale dependence of $b(R)$ because local
gas temperature is related to the gravitational potential, and thus
contains information about the large scale density field. We show that
temperature dependence generally causes $b(R)$ to vary on quasilinear
and nonlinear scales, indicating that scale dependence of bias may be
a generic effect in realistic galaxy formation scenarios.  We find
that the relationship between the galaxy and mass density fields is
also a function of galaxy age. On large scales, the older galaxies are
highly biased ($b\approx1.7$) and highly correlated
($r\equiv\avg{\delta\delta_g}/\sigma\sigma_g\approx1.0$) with the mass
density field; younger galaxies are not biased ($b\approx0.8$) and are
poorly correlated ($r\approx0.5$) with the mass. We argue that linear
bias is inadequate to describe the relationship between galaxies and
mass. We present a more physically based prescription which better fits
our results and reproduces the scale dependence of the bias:
$\rho_g/\avg{\rho_g} = L (\rho/\avg{\rho})^M (1+T/40,000
{\rm ~K})^N$, where $L=1.23$, $M=1.9$, and $N=-0.66$.

\end{abstract}


\section{Motivation}
\label{motiv}

Imminent large-scale galaxy redshift surveys such as the Sloan Digital
Sky Survey (SDSS; \cite{gunn95ap}) and the Two-Degree Field (2DF;
\cite{colless98ap}) will probe the galaxy density field of the
universe with unprecedented precision. Were galaxies accurate tracers
of the mass density field, the results of these surveys would put
severe constraints on cosmological models. If the Cold Dark Matter
(CDM) picture for the linear theory power spectrum is correct, these
surveys can in fact measure cosmological parameters, such as the mass
density $\Omega_0$, the vacuum density $\Lambda$, and the baryon
density $\Omega_b$ (\cite{delaix98ap}; \cite{tegmark98bp};
\cite{goldberg98ap}; \cite{wang99ap}). However, the visible matter in
galaxies is only a small percentage of the baryons in the universe,
which in turn is a small percentage of the mass in the universe
(\cite{fukugita98ap}; \cite{cen98bp}). Moreover, the process of galaxy
formation is complex and nonlinear, including a complicated interplay
between gravitational fields, hydrodynamics, microphysics and star
formation. Does this complicated process produce a population of
galaxies whose number density field traces the mass density field
perfectly? In this introductory section, we argue that observations
already suggest that it does not: that the relationship between the
density fields of galaxies and mass is biased, scale dependent, and
nonlinear, as well as dependent on morphological type. This discussion
provides the observational motivation for asking the theoretical
question: how is the galaxy density field related to that of the mass?

The crucial element of our argument is that different morphological
types of galaxies have different density fields (\cite{hubble36ap};
\cite{oemler74ap}).  Consider the observed differences between the
clustering strengths of galaxies of different types. Various authors
have compared elliptical and spiral galaxies, generally finding that
the fluctuation amplitude of ellipticals is stronger than that of
spirals by a factor of 1.3--1.5 (\cite{davis76ap};
\cite{giovanelli86ap}; \cite{santiago92ap};
\cite{loveday96ap}; \cite{hermit96ap}; \cite{guzzo97ap}).  Similarly, a comparison of
the galaxy distribution in the \iras\ redshift survey
(\cite{strauss92bp}) with those in the Center for Astrophysics
Redshift Survey (CfA; \cite{huchra83ap}) and in the Optical Redshift
Survey (ORS; \cite{santiago95ap}), shows that optically-selected
galaxies are clustered more strongly than infrared-selected galaxies
by a similar factor (\cite{davis88ap}; \cite{babul90ap};
\cite{strauss92ap}).  These differences in the {\it
amplitude} of clustering may be accounted for by invoking a
deterministic linear bias prescription:
\begin{equation}
\label{linbias}
\delta_g(\vv{r}) = b\, \delta(\vv{r})\mathrm{,}
\end{equation}
where $\delta_g(\vv{r}) \equiv \rho_g(\vv{r})/\avg{\rho_g} -1$ is the
galaxy overdensity and $\delta(\vv{r}) \equiv \rho(\vv{r})/\avg{\rho}
- 1$ is the mass overdensity, smoothed on some scale. To explain the
observations, one must assume that different galaxy populations are
``biased'' by different factors $b$; clearly all but one of these
bias factors must differ from unity.  Furthermore, there are
differences between the shapes of correlation functions of galaxies of
different types, at least at small scales. For instance, the
ratio of the correlation functions of ellipticals and
spirals found by \cite{hermit96a} and \cite{guzzo97a} declines
with scale over the range between 1 and 10 $h^{-1}$ Mpc.  This scale
dependence cannot result from a deterministic linear bias. Therefore,
there must exist a more complicated relation between the density
fields of different morphological types.

Alternatively, consider the density-morphology relation, quantified by
\cite{dressler80a}, \cite{postman84a}, and \cite{whitmore93a}. In the
field, spirals comprise about 70\% of all galaxies, and ellipticals
and lenticulars comprise the rest; in the cores of rich clusters the
situation is reversed, and ellipticals and lenticulars account for
90\% of all galaxies. The relationship between spiral density and the
density of all galaxies is extremely nonlinear in the densest cluster
regions. 

These differences among different morphological types suggest that the
relationship between all galaxies and mass is comparably
complicated. After all, it would be a coincidence if the overdensity
field of all galaxies exactly traced the full mass overdensity,
despite the fact that the different morphologies have formed at
different times and with different efficiencies. In any case, the
selection effects of redshift surveys (color, surface brightness,
luminosity, {\it etc.})  will cause any catalog to contain a mix of
morphological types that differs from the mix in a volume-limited
sample.  Since the overdensity fields of different morphologies have
different density fields, the results of every survey are ``biased''
to some degree. It is thus interesting to explore theoretically how
the mass density in the universe might be related to the galaxy
density, and to the density of different morphological types.

One approach is to express the galaxy density as a local
transformation of the dark matter density (or other variables), the
simplest version of which is the deterministic linear bias of Equation
(\ref{linbias}). Peaks-biasing (\cite{bardeen86ap}) and threshold
biasing (\cite{kaiser84ap}) were the first suggestions along these
lines. Indeed, \cite{davis85a} used the peaks-biasing scheme to
reconcile an $\Omega_0=1$ universe with observations. \cite{fry93a},
\cite{coles93a}, \cite{mann98a}, and \cite{narayanan98a} have studied
how nonlinearity and other complications in this transformation can
affect statistics such as the density distribution function, the
correlation function, and $M/L$ ratios of clusters.  Another
possibility is that the local transformation is not deterministic;
\cite{scherrer98a} show how such a stochastic relation affects the
correlation function and the power spectrum of galaxies.
\cite{dekel98a} have developed a general formalism (which we adopt
here) for expressing such a nonlinear and stochastic relation.

A second approach is to use semi-analytic models to simulate galaxy
formation ({\it e.g.}, \cite{white87ap}; \cite{white91ap};
\cite{cole94ap}; Kauffmann {\it et al.}~1993, 1998;
\cite{somerville98ap}). These methods use either
$N$-body simulations or the formalism of \cite{press74a} and
\cite{mo96a} to follow dark matter halos and their merging
histories. Simple rules for star formation and feedback follow the
evolution of galaxies inside these halos. \cite{kauffmann97a} and
\cite{kauffmann98a} used such models to study the relation between
mass density and galaxy density. We compare their
results to ours below.

The third approach, which we adopt here, is to use hydrodynamic
simulations with heuristic models of galaxy formation built into
them. Generically, these simulations follow the radiative physics of
the gas and use physically motivated prescriptions to convert baryonic
fluid into collisionless stellar particles, which is necessary due to
our ignorance of the details of the star-formation processes and the
finite resolution of the code.  \cite{carlberg89a}, \cite{evrard94a},
and Katz, Hernquist, \& Weinberg~(1992,~1998) have all used the
Lagrangian smoothed particle hydrodynamic (SPH) method to address the
question of galaxy bias; Gnedin (1996a,b), Cen \&
Ostriker~(1992,~1998a), and this paper use Eulerian methods for the
same purpose.

In this paper, we examine the galaxy density field produced by the
large-scale hydrodynamic simulations of \cite{cen98a}.  With recent
improvements in computational methods and in computer hardware,
the dynamic range in these simulations is approaching the level at
which one can examine the formation of galaxies in a cosmological
context. In Section \ref{formalism}, we present the formalism of
\cite{dekel98a} for expressing the relation between galaxies and mass,
and review some previous theoretical results. In Section
\ref{simulations}, we present details of the numerical simulations of
\cite{cen98a}. In Section \ref{onevar}, we examine the relation
between the distributions of galaxies and mass in these simulations,
as a function of smoothing scale and age. In Section \ref{twovar}, we
show that one can explain the properties of the galaxy density field
more completely by allowing for its dependence on gas temperature as
well as on mass density.  We present an analytic fit to our results
which describes the galaxy density field well. In Section \ref{toy},
we show that the dependence on gas temperature explains the scale
dependence of the bias. Finally, we introduce a toy model which
reproduces some of the salient properties of the galaxy density field,
and shows how scale dependence could be a generic property of galaxy
formation. We discuss some directions for future work in Section
\ref{discussion}.

\section{Formalism}
\label{formalism}

In this section, we present a general formalism developed by
\cite{dekel98a} for expressing the present-day Eulerian relation
between the galaxy and mass density fields smoothed on a local scale
$R_0$.  Such an approach, which considers only the density fields at
$z=0$, of course does not account for the fact that two regions which
have similar properties now may have had significantly different
histories. We relate this formalism to the simple approximation that
the galaxy and mass densities form a bivariate Gaussian distribution.
Finally, we describe how certain complications allowed by this
formalism, namely nonlinearity and stochasticity, can cause the
relation between the two density fields to be a function of
scale. This formalism is not a theoretical model in its own right, but
rather gives a useful framework for quantifying all relevant aspects
of galaxy bias.

We begin by defining $P(\delta_g | X_i)$ as the conditional
probability of a certain overdensity of galaxies $\delta_g \equiv
\rho_g/\avg{\rho_g}-1$, given the set of {\it local} conditions
expressed by $X_i$, where the $X_i$ can be mass density, temperature,
velocity shear, or any other present-day parameter thought to be
relevant to galaxy formation. We use this probabilistic formalism
despite the fact that galaxy formation is a deterministic process, for
two reasons. First, two regions with similar properties at $z=0$
may have had quite different histories; these differences may make it
impossible to find a perfectly deterministic relation between
$\delta_g$ and other Eulerian variables. Second, even if such a
relation did exist, one might not be able to identify the correct set
of Eulerian variables with which to express it. Choosing a wrong or
incomplete set of $X_i$ would cause the relation to have large
scatter.  Because this resulting stochasticity would in fact have a
physical basis, it would be likely to have interesting statistical
properties, such as spatial correlations. Thus, one of our goals will
be to find a set of $X_i$ which minimizes this scatter.

Having defined the conditional probability, one can define the
conditional mean:
\begin{equation} 
\avg{\delta_g| X_i} \equiv \int d\delta_g\,\delta_g P(\delta_g|X_i)
\mathrm{.}
\end{equation} 
The variance of the scatter about this mean is:
\begin{equation}
\sigma_b^2 \equiv \avg{\epsilon^2} = \avg{(\delta_g -
\avg{\delta_g|X_i})^2}
\mathrm{,}
\end{equation}
where, as indicated, $\epsilon$ represents the residuals of the
conditional mean galaxy density at scale $R_0$. The quantity
$\sigma_b/\sigma_g$, where $\sigma_g^2\equiv\avg{\delta_g^2}$,
expresses the degree of stochasticity in the relation between galaxies
and the variables $X_i$.  Typically, investigators in this field have
assumed that the most important (if not the only) local condition
worth considering is the mass overdensity $X_i = \delta \equiv
\rho/\avg{\rho}-1$ (an exception being \cite{narayanan98ap}). In this
case, one considers the conditional probability $P(\delta_g|\delta)$
and the corresponding conditional mean $\avg{\delta_g|\delta}$, which
is meant to summarize the relation between galaxies and mass on the
given scale $R_0$.

If the joint distribution of $\delta$ and $\delta_g$ is a bivariate
Gaussian at scale $R_0$, an appropriate approximation in the linear
regime, the joint probability can be expressed as:
\begin{equation}
P(\delta,\delta_g) = \frac{1}{2\pi\sqrt{\sigma^2 \sigma_g^2 -
\langle\df\df_g\rangle^2}} \exp\left(-\frac{1}{2}
\frac{\sigma_g^2\df^2 - 2 \langle\df\df_g\rangle \df\df_g + \sigma^2
\delta_g^2}{\sigma^2 \sigma_g^2 - \langle\df\df_g\rangle^2}\right)
\mathrm{.}
\end{equation}
We will refer to this special case as ``linear bias.'' Such a
distribution can be completely characterized by three quantities:
$\sigma^2\equiv\avg{\delta^2}$, $b\equiv\sigma_g/\sigma$, and
$r\equiv\avg{\delta\delta_g}/\sigma\sigma_g$, and indeed, this
motivates the definitions of these quantities for arbitrary $P(\delta,
\delta_g)$.  The quantity $b$, hereafter, the ``bias,'' compares the
r.m.s.~amplitude of the galaxy overdensity to that of the mass
overdensity. The quantity $r$, hereafter the ``correlation
coefficient,'' expresses how closely the galaxy density field traces
the mass density field. That is, if $r=\pm 1$, the relation between
mass and galaxies is deterministic; if $r=0$ the galaxies are
distributed independently of the mass. Note that the Gaussian
assumption implies:
\begin{eqnarray}
\avg{\delta_g|\delta} &=& b\, r\, \delta \mathrm{;}\cr
\sigma_b/\sigma_g &=& \sqrt{1-r^2} \mathrm{\quad (Gaussian~model).}
\end{eqnarray}

In the nonlinear regime, the density field is far from Gaussian;
however, calculating the second moments $b$ and $r$ will still give
useful information on how the galaxy and mass density fields relate.
The quantity $br$ for any given pair of galaxy and
mass density fields is the slope of a linear regression
of $\delta_g$ on $\delta$. Similarly, $b/r$ is the slope of the linear regression
of $\delta$ on $\delta_g$. The quantity $\sqrt{1-r^2}$ is a measure of
the scatter around either regression. The scatter can occur either
because nonlinearities make a straight line a poor approximation to
$\avg{\delta_g|\delta}$, or because of stochasticity. As
\cite{dekel98a} note, the ratio $\sigma_b/(\sigma_g\sqrt{1-r^2})$
measures the contribution of stochasticity (as opposed to
nonlinearity) to the total scatter around the linear regression $br$.

In the context of the formalism presented here, let us examine how
scale dependence in the relation between galaxies and mass may arise.
Most work to date assumes that the deterministic linear bias of
Equation (\ref{linbias}) holds ({\it i.e.} $r=1$). Such a relation is
scale independent; the same factor $b$ applies at all scales. However,
$\avg{\delta_g|\delta}$ may in general be a nonlinear function of
$\delta$. In fact, on scales at which $\sigma \gg 1$, this
nonlinearity inevitably results from the condition $\delta_g > -1$ (as
long as $\avg{\delta_g|\delta}$ does not {\it exactly} equal
$\delta$). A simple approach is to expand $\delta_g$ in a Taylor
series around $\delta$ (\cite{fry93ap}):
\begin{equation} 
\label{taylor}
\avg{\delta_g|\delta} = b_1 \delta + \frac{b_2}{2} (\delta^2 -
\avg{\delta^2}) +
\frac{b_3}{6} (\delta^3 - \avg{\delta^3}) + \ldots
\end{equation} 
The introduction of nonlinearity opens the door to scale dependence;
if $\delta_g$ and $\delta$ are smoothed on a scale above $R_0$, the
coefficients in Equation (\ref{taylor}) may change. On the other hand,
\cite{scherrer98a} show that for hierarchical clustering, even in the
presence of nonlinear bias on small scales, $b(R)$ is independent of
$R$ on large scales.

In addition to being nonlinear, the relation can be stochastic, such that
$\sigma_b\ne 0$. \cite{scherrer98a} have found that $b$ is independent of scale
in this case as well.  However, they assumed that the residual field $\epsilon$
about $\avg{\delta_g|\delta}$ is spatially uncorrelated. However, the scatter
$\epsilon$ may have a physical basis; thus, it is possible that $\epsilon$
correlates with the large scale density field. If such a correlation existed,
$b(R)$ would be a function of scale. To illustrate this possibility, assume for
the moment that at some small smoothing scale $R_0$, the joint
$\delta_g$-$\delta$ distribution is a bivariate Gaussian (Equation 4), with a
bias of $b(R_0)$ and correlation coefficient $r(R_0)$. We define $\epsilon
\equiv \delta_g - br\delta$ on this scale. We can subsequently smooth
$\delta_g$, $\delta$, and $\epsilon$ over a large scale $R\gg R_0$. In this
case:
\begin{equation}
\label{bvar}
b^2(R) = b^2(R_0) r^2(R_0) + \frac{\avg{\epsilon^2}_R 
+ 2b(R_0)r(R_0)\avg{\delta\epsilon}_R} {\sigma^2(R)}  \mathrm{.}
\end{equation}
As $R\rightarrow R_0$, by definition we have
$\avg{\delta\epsilon}_R\approx 0$ and $\avg{\epsilon^2}_R \approx
\sigma_g^2(R_0)(1-r^2(R_0))$, so that $b(R\rightarrow R_0)=b(R_0)$ as
necessary. 
If $\avg{\delta\epsilon}_R$ or $\avg{\epsilon^2}_R$ varies
on larger scales, clearly $b(R)$ will vary as well.  We show
below that this variation can result from fairly simple physical
considerations; indeed, the effects are strong in the simulations
considered here. 

In the following sections, we will describe the numerical simulations
of \cite{cen98a} and present the results in the context of the
formalism presented here. As the independent variable $X_i$, we at
first use the traditional dark matter overdensity $\delta$, and find
that the dependence on $\delta$ cannot completely characterize the
galaxy density field. We will
find that using the description $P(\delta_g|\delta,T)$, where $T$ is
the local gas temperature, gives a much more satisfactory description
of the galaxy density field. In particular, accounting for the
dependence on $T$ (or its counterpart, the dark matter velocity
dispersion $\avg{v^2}$) also accounts for most of the scale dependence
of the bias. Thus, temperature dependence causes stochasticity in the
galaxy-mass relation, and this stochasticity is correlated with the
mass distribution over large scales, causing $b(R)$ to vary with
scale.

\section{Simulations}
\label{simulations}

For these simulations, the work of \cite{ostriker95a} motivated the
choice of a flat cold dark matter cosmology with $\Omega_0 = 0.37$,
$\Omega_\Lambda = 0.63$, and $\Omega_b=0.049$. Recent observations of
high-redshift supernovae have lent support to the picture of a flat,
low-density universe, though great uncertainty still remains
(\cite{perlmutter97ap}; \cite{garnavich98ap}). The
Hubble constant was set to $H_0 = 100\,h$ km s$^{-1}$ Mpc$^{-1}$, with
$h=0.7$.  The primordial perturbations were adiabatic and random
phase, with a power spectrum slope of $n = 0.95$ and amplitudes such
that $\sigma_8=0.8$ for the dark matter at $z=0$, at which time the
age of the universe is 12.7 Gyrs. We use a periodic box 100 $h^{-1}$
Mpc on a side, with $512^3$ grid cells and $256^3$ dark matter
particles. Thus, the mass resolution is about $5\times 10^9 M_\odot$
and the grid cell size is $\sim$ 200 $h^{-1}$ kpc.  The smallest
smoothing length we consider is a 1 $h^{-1}$ Mpc radius top hat, which
is considerably larger than a cell size. On these scales and larger,
the relevant gravitational and hydrodynamical physics are correctly
handled.  On the other hand, subgrid effects such as the fine grain
structure of the gas and star formation may influence large-scale
properties of the galaxy distribution. As we will decribe, we handle
these effects using plausible, though crude, rules.

\cite{cen98a} describe the hydrodynamic code in detail; it is similar
to but greatly improved over that of \cite{cen92a}.  The simulations
are Eulerian on a Cartesian grid and use the Total Variation
Diminishing method with a shock-capturing scheme.  In addition, the
code accounts for cooling processes and incorporates a heuristic
galaxy formation criterion, whose essence is as follows: if a cell's
density is high enough, if the cooling time of the gas in it is
shorter than its dynamical time, if it contains greater than the Jeans
mass, and if the flow around that cell is converging, it will have 
stars forming inside of it. The code turns a fraction of the baryonic
fluid component into collisionless stellar particles (hereafter
``galaxy particles''), which subsequently contribute to metal
production and the background ionizing UV radiation.  The masses of
these galaxy particles range from about $10^6$ to $10^9$
$M_\odot$. Thus, many galaxy particles are contained in what would
correspond to a single galaxy in the real universe. Instead of
grouping the particles into galaxies, we simply define a galaxy mass
density field from the distribution of galaxy particles themselves.
Thus our results will {\em not} be directly comparable to observations
of the density field based on galaxy counts. 

Ideally, we would like to study the properties of $\delta_g$ for different
galaxy morphologies. Our simulations clearly do not have the resolution
necessary to determine the morphology of individual galaxies based on their
internal structure. However, there is a rough correlation between a galaxy's
morphology and its star-formation history (\cite{roberts94ap};
\cite{kennicutt98ap}).  Keeping this in mind, we can examine the simulations
for an age-density relation in analogy to the morphology-density relation of
\cite{dressler80a} and \cite{postman84a}.  \cite{cen93a} did so, finding
qualitatively that the oldest galaxies had fallen into clusters and that the
younger galaxies were forming in the lower density regions. In addition,
\cite{cen98a} have examined the dependence of the power spectrum on galaxy
age. Here, we visit the problem again, now looking in more detail at the joint
distribution of the mass density and the density of each galaxy
population. We 
split the galaxy particles into four age quartiles, defined such that the total
mass of galaxy particles in each quartile is the same. These age quartiles are
not meant to be taken as literally corresponding to different morphologies,
since there are other variables besides age which determine galaxy type;
however, their differences should at least be indicative of the differences
between different galaxy morphologies. The age and redshift ranges of the
quartiles are given in Table~\ref{quartiles}.  As we examine the properties of
the full galaxy mass density field, we do the same for the density fields of
each age quartile.

\begin{figure}
\figurenum{\fignum a}
\epsscale{1.0}
\plotone{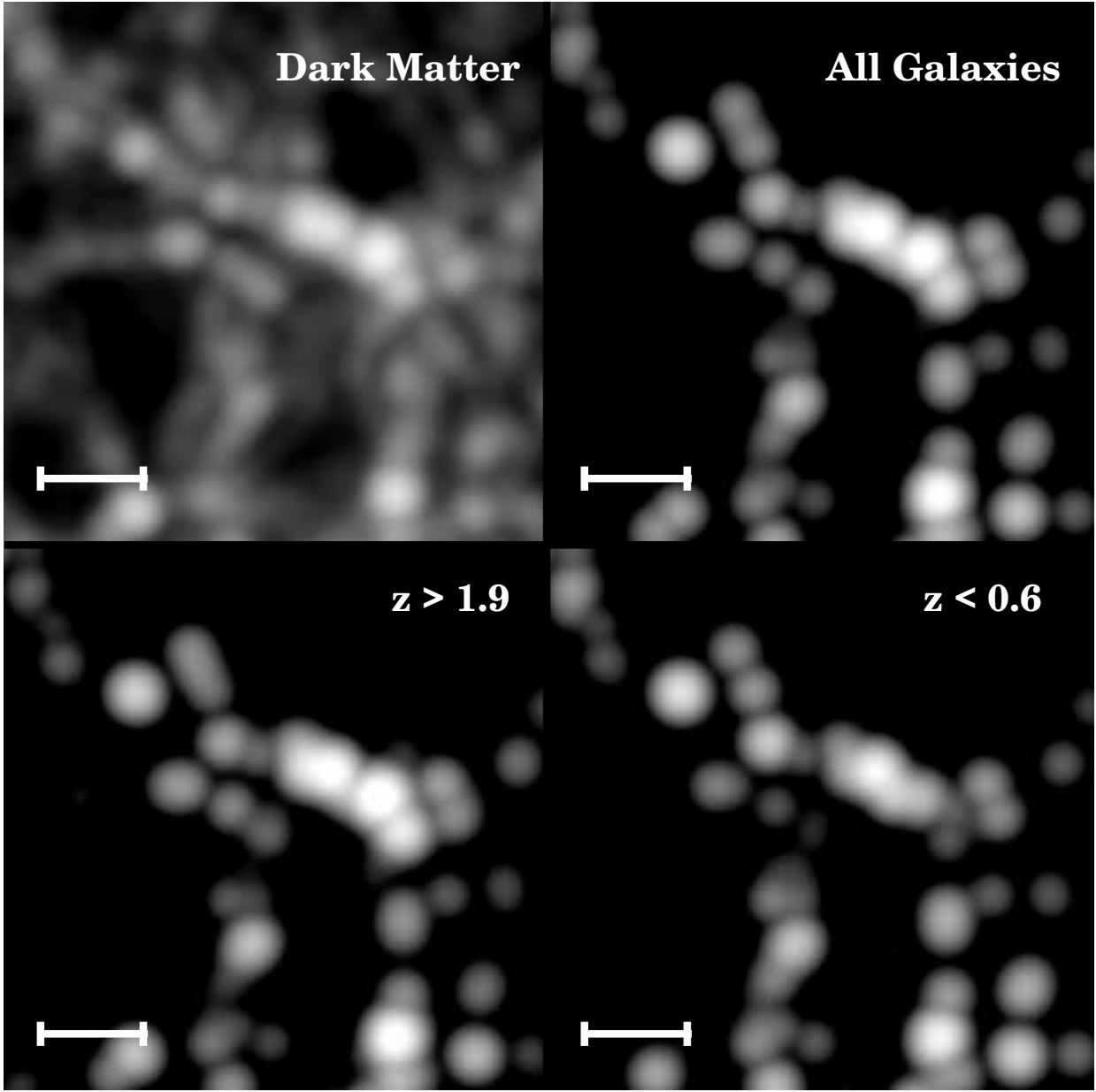}
\epsscale{1.0}
\caption{\label{quad.ss} A slice through our simulation 50 $h^{-1}$
Mpc on a side (one half the total box length). As labeled, the
quadrants show the fractional overdensity $\delta$ in dark matter, in the galaxies, in the
young galaxies, and in the old galaxies, in clockwise order. The
stretch is logarithmic and is set the same in all quadrants. The
fields are smoothed with a 1 $h^{-1}$ Mpc Gaussian filter. The scale
bar
indicates 10 $h^{-1}$ Mpc. Note the large voids in the galaxy
distribution and the reduction in the fraction of young galaxies in
the large overdensity near the center.}
\end{figure}

Figures \ref{quad.ss} and \ref{quad.ls} show a slice through the
galaxy and dark matter density fields 50 $h^{-1}$ Mpc on a side, at 1
$h^{-1}$ Mpc and 10 $h^{-1}$ Mpc smoothing, respectively. From top
left, in clockwise order, we show the quantity $\delta$ for the dark
matter, all of the galaxy 
particles, the youngest galaxy quartile, and the oldest galaxy
quartile. For each of these, $\delta$ is normalized to the mean
density of the sample in question, so that in the absence of biasing,
these plots would be identical.  Evidently the galaxy distribution
follows the dark matter 
distribution well, except in the underdense regions, which are
completely empty voids in the galaxy particle distribution. It is
apparent from these pictures that the youngest galaxy particles are
distributed quite differently than is the mass. At small smoothing scales
the effect is obvious only in the clusters; otherwise the young
galaxies follow the dark matter. At large smoothing scales, however,
the young galaxies are underdense in the clusters, and their density
fields peak along the filaments. On the other hand, the oldest
galaxies follow the mass distribution well on all scales, and are
quite obviously biased. In the rest of the paper, we will quantify the
differences among these various density fields.

\section{Single-Variable Bias}
\label{onevar}

In this section, we study the relation between the galaxy and mass
density fields, expressed by $P(\delta_g|\delta)$ and
$\avg{\delta_g|\delta}$. We begin by smoothing the galaxy density
field over several different scales and showing that the relationship
between galaxies and mass is a function of scale. Then we show that
it is the properties of the scatter about $\avg{\delta_g|\delta}$ at
small scales, and not the form of $\avg{\delta_g|\delta}$ itself,
which causes this scale dependence.

\subsection{Galaxy Density vs. Mass Density}

First we directly compare the density field of dark matter to that of
galaxies. We do so by plotting in Figure~\ref{alldd} the conditional
probability $P(1+\delta_g|1+\delta)$ using top hat smoothing filters
of six different radii: 1, 2, 5, 8, 16, and 30 $h^{-1}$ Mpc. We use
$1+\delta$ here simply for convenience in plotting the results. The
greyscale in this figure is a logarithmic stretch of
$P(1+\delta_g|1+\delta)$; the greyscale for each column is normalized
separately.  Note that for small smoothing scales, the discreteness of
the dark matter particles limits our measurement of $\delta$ in voids.
The vertical dashed line is the density corresponding to about 50
particles within one smoothing length, below which this effect becomes
important.  Note further that there is structure in the histograms at
large smoothing scales, since there are many bins in the histogram but
only a few truly independent values in the periodic volume.  We plot
$\avg{\delta_g|\delta}$ for each smoothing scale as the solid black
lines; the 1$\sigma$ deviations from this mean line are shown as the
dotted black lines. It is immediately apparent that this function is
nonlinear, and that there is large scatter about it.

The same comparison can be made for the galaxies in each
age quartile separately, as is shown in Figures~\ref{agedda} and \ref{ageddb}
for top hat smoothing filters of two radii: 1 and 30 $h^{-1}$ Mpc. The
overdensity $\delta_g$ for each quartile is defined by normalizing to the mean
density of that quartile. Note that there is a tight and highly biased
relation between the distributions of older galaxies and the dark matter. On the other
hand, the relation between the youngest galaxies and the mass is quite
stochastic, even at the largest scales. At small scales,
$\avg{\delta_g|\delta}$ for young galaxies is not monotonic; as in the
real universe, young galaxies rarely live in the highest density
regions. This trend with age is easily understood.  The densest
regions of the universe are in the deepest potential wells, so they
have the hottest gas with the longest cooling times; thus, this gas
stopped forming galaxies some time ago.  As a consequence, on average,
the galaxies in the densest regions are older than galaxies elsewhere.

One can quantify the relation between galaxies and mass by calculating
second moments of the galaxy-mass distribution. Thus, at each smoothing
scale, we calculate the parameters $b\equiv\sigma_g/\sigma$ and $r \equiv
\avg{\delta_g\delta}/\sigma_g\sigma$ and list their values in Table
\ref{momtable}. Note that $b$ declines strongly with scale, from 2.6
at 1~$h^{-1}$~Mpc to 1.2 at 30~$h^{-1}$ Mpc; this behavior is
consistent with the work of \cite{cen92b}.  Meanwhile, $r \sim 0.9$
almost independent of scale, meaning that galaxies are well-correlated
with mass. Similarly, we can calculate $\sigma_b$, the variance of the
scatter about $\avg{\delta_g|\delta}$, and find that
$\sigma_b/\sigma_g \sim 0.3$--$0.4$. For comparison to $r$, we list
the ratio $\sigma_b/\sigma_g\sqrt{1-r^2}$, which is $\sim 0.8$--$0.9$
at all scales, indicating that stochasticity, rather than
nonlinearity, dominates the scatter around the linear regression
slopes $br$ and $b/r$.

The parameters $b$ and $r$ will also clearly depend on what age
galaxies one considers, as Figures~\ref{agedda} and \ref{ageddb}
indicate. In Table \ref{agetable} we list $b$ and $r$ for each of the
quartiles and smoothing scales used in Figures~\ref{agedda} and
\ref{ageddb}. Again, the dependence on scale is evident. However, more
prominent are the differences between galaxies of different
ages. Older galaxies are much more highly biased than young galaxies
at all scales. In addition, older galaxies are more correlated with
the mass distribution; $r \sim 0.9$--$1.0$ for the oldest galaxies,
while $r\sim 0.5$ for the youngest galaxies. These results are
agree with
observations that early-type galaxies are biased relative to late-type
galaxies.  The low correlation coefficient for young galaxy particles
means that fluctuations in their density field are poorly correlated
with fluctuations in the mass density field, as one can see from
Figures \ref{quad.ss} and especially \ref{quad.ls}; that $b$ is near unity
on large scales merely indicates that the fluctuations in the two
fields are of similar amplitudes.

\subsection{Scale Dependence}

To study more carefully the scale dependence of the bias, we show $b(R)$ in
Figure \ref{b} and $r(R)$ in Figure \ref{r}.  The solid black lines refer to
all the galaxies; the dashed lines refer to each age quartile, as
labeled. These curves show the same behavior found in Tables~\ref{momtable} and
\ref{agetable}.  The older galaxies are more biased than the younger galaxies.
Furthermore, the density field of the oldest quartile is extremely
well-correlated with the mass density field, while that of the youngest
quartile is extremely poorly correlated.  Meanwhile, bias declines with scale
for all the galaxies and for each quartile.  Other investigators have published
comparable results, and it is appropriate here to address the similarities and
differences between the current results and those of others. We first note that
these results are qualitatively similar to the previous results of
\cite{cen92b} using the same method but with different cosmologies. We also,
for the purposes of this discussion, define $b_\xi(R)\equiv \xi_g(R)/\xi(R)$,
the ratio of the galaxy and mass correlation functions on scale $R$. It is
roughly, though not exactly, comparable to our definition
$b\equiv\sigma_g/\sigma$.

The most recent $N$-body results, using the Adaptive Refinement Tree
(ART) method (\cite{kravtsov97ap}), indicate, in fact, that there is
antibias at small scales, and that the bias increases with scale
(\cite{kravtsov98ap}). This effect seems to be due to the merging and
destruction of halos in dense regions. \cite{klypin99ap} claim that
ART has sufficient dynamic range to avoid the ``overmerging'' problem
that has plagued pure $N$-body simulations (\cite{white76ap};
\cite{frenk88ap}; \cite{vankampen95ap}; \cite{summers95ap};
\cite{moore96ap}). Since we track stellar mass density, rather than
galaxy number density, we cannot address the questions of mergers and
destruction, and thus these effects are not apparent in our
results. 

On the other hand, the results of the SPH simulations of \cite{carlberg89a},
\cite{katz92a}, and \cite{evrard94a} all show a slight dependence of $b_\xi(R)$
on $R$ between 1 and 10 $h^{-1}$ Mpc, and a substantial increase of
the bias on smaller scales ($\sim 0.3$ $h^{-1}$ Mpc). These results are in
qualitative agreement with ours, and it is likely that the physical effects
(described in Section \ref{toy}) which cause the scale-dependence in our
simulations are also important in the SPH simulations, since the criteria for
producing galaxies in those simulations are similar to those in ours.

\cite{kauffmann97a} used semi-analytic modelling techniques combined
with $N$-body simulations to explore the relationship between the
clustering of galaxies and mass. Their method has the advantage that
it can efficiently explore parameter space; however, it cannot model
the gas dynamics or the effects of environment. For galaxies with
$M_B<-20$ in an open $\Omega_0=0.2$ CDM model, they find a similar
scale dependence of the bias to ours (their Figure 6; note that their
definition of ``bias'' is close to $br$, the regression of galaxy
density on mass density, in our notation). For the $\Omega_0=1$ CDM
model, they find that the scale-dependence is present but less
acute. On the other hand, \cite{kauffmann98a} shows $b_\xi\equiv
\xi_g(r)/\xi(r)<1$ on small scales and increasing to larger scales for
their $\Lambda$CDM cosmology, and $b$ slightly greater than one for
the $\tau$CDM cosmology. \cite{kauffmann98a} normalize their galaxy
formation parameters somewhat differently than do \cite{kauffmann97a},
such that galaxies of the same luminosity are in lower mass halos in
the later work; this accounts for part of the difference between their
results.  In sum, these semi-analytic models, which have a vastly
different model of galaxy formation from ours, can in some cases
produce scale-dependence similar to that found here. 

We are interested in discovering exactly why bias decreases with scale in our
results. Scale dependence can be due to two things only: the nonlinearity of
the locally defined $\avg{\delta_g|\delta}$, or the properties of the field of
residuals $\epsilon$ about that mean.  Several theoretical forays suggest that
the first possibility cannot be the case (\cite{coles93ap};
\cite{scherrer98ap}; \cite{dekel98ap}). We can test these results by applying
the $\avg{\delta_g|\delta}$ defined on small scales to the mass density
field. Specifically, we calculate $\avg{\delta_g|\delta}$ for the galaxy and
mass fields smoothed with a 1 $h^{-1}$ Mpc radius top hat; then, at each grid
cell, we check the value of $\delta$ and set the galaxy density in that cell to
the appropriate value of $\avg{\delta_g|\delta}$.  We refer to the resulting
field as the ``fake galaxy'' density field.

We can now consider the statistic $b(R)$ for this fake galaxy field,
which we determined using the mass density field alone, shown as the
dotted lines in each panel of Figure~\ref{bth}. Here, $R$ takes into
account the 1 $h^{-1}$ Mpc smoothing already present in $\delta$.  The
solid lines are the actual $b(R)$ for the galaxies found in the
simulations, from Figure~\ref{b}. The top panel corresponds to all the
galaxies, the middle panel to the oldest galaxies, and the bottom
panel to the youngest galaxies.  This procedure does not reproduce the
behavior of bias as a function of scale, at least when we use this
single-variable model $\avg{\delta_g|\delta}$.  There is some scale
dependence, which is allowed by the results of \cite{scherrer98a} in
the nonlinear regime.  At small scales, the fake galaxy distribution
has less power than the actual galaxy distribution, while at large
scales, the fake galaxy distribution remains much more biased than the
actual galaxy distribution.  Thus, the mean relation
$\avg{\delta_g|\delta}$ between galaxies and mass at small scales does
not contain all the information about $\delta_g$.

The remaining possibility is that the properties of the residual field
$\epsilon$ about $\avg{\delta_g|\delta}$ cause the scale dependence.
These residuals must correlate over large scales in such a way that
$\epsilon<0$ in large scale overdensities and $\epsilon>0$ in large
scale voids. In that case, $\avg{\delta\epsilon}_R$ will be negative
on large scales and, as Equation (\ref{bvar}) reveals, $b(R)$ will
decrease with scale. To test this possibility, we calculate $\epsilon$
at 1 $h^{-1}$ Mpc top hat smoothing, and then smooth $\epsilon$ and
$\delta$ on scale $R$ in order to calculate
$\avg{\delta\epsilon}_R/\sigma^2(R)$.  Figure \ref{epsdel} shows the
behavior of this quantity. As one would predict, it is near zero when
$R \sim 1$ $h^{-1}$ Mpc, and becomes strongly negative on larger
scales. The same holds for the youngest and oldest quartiles, also
shown in Figure~\ref{epsdel}. This result indicates that, indeed, the
residual field tends to be negative in large scale overdensities, and
positive in large scale voids.

In other words, the scatter about $\avg{\delta_g|\delta}$ has
interesting statistical correlations. It depends on other variables
which are important to galaxy formation. The dependence on and the
nature of these variables must be such as to reproduce the scale
dependence found in this section. In the next two sections, we
investigate the dependence of galaxy density on a number of other
variables, and find that accounting for the dependence of galaxy
density on local temperature can both reduce the scatter in the
residual field and explain the scale dependence of $b$.

\section{Two-Variable Bias}
\label{twovar}

From the last section we learned that modeling the galaxy density
field as a function of mass density alone was unsatisfactory. That is,
spatial correlations in the scatter $\epsilon$ are important. We would
like to both reduce the scatter in our estimate of the galaxy density
and to explain the scale dependence of $b(R)$. To do so, we can add an
independent variable to our conditional probability and consider the
function $P(\delta_g | \delta, X_i)$. Note that to account for the
scale dependence, $X_i$ must have spatial correlations similar to
those of $\epsilon$; we discuss this point further in the next
section.  We try several $X_i$ and find that the most successful is
the local gas temperature $T$, or, equivalently, the local dark matter
velocity dispersion. This result is perhaps not surprising, since gas
temperature is surely an important parameter in galaxy formation.
That is, in order to form stars, gas must be able to cool efficiently
and collapse, which cannot happen at very high temperatures. In
addition, we will find in Section \ref{toy} that accounting for this
dependence on $T$ also accounts for the dependence of $b(R)$ on scale.

For each $X_i$ we choose as a second independent variable, we want to know
how much it reduces the stochasticity in the relation $P(\delta_g | \delta,
X_i)$. To quantify this, we  calculate the ratio:
\begin{equation}
\frac{\sigma_{b,2}}{\sigma_b} \equiv \frac{\Avg{(\delta_g - \avg{\delta_g |
\delta, X_i})^2}^{1/2}}{\Avg{(\delta_g - \avg{\delta_g |
\delta})^2}^{1/2}} \mathrm{,}
\end{equation}
which expresses the scatter of $\delta_g$ around
$\avg{\delta_g|\delta,X_i}$, compared with the scatter around
$\avg{\delta_g|\delta}$.  Thus, if $X_i$ is perfectly correlated with
$\delta$, or if it is perfectly uncorrelated with $\delta_g$, then
$\sigma_{b,2}/\sigma_b \approx 1$. In the next few paragraphs, we will
substitute for the variable $X_i$ the following: $T$, the local
temperature; $\avg{v^2}$, the local dark matter velocity dispersion;
the prolateness and oblateness of the local dark matter density field;
and the shear in the dark matter velocity field.  Our results
for $\sigma_{b,2}/\sigma_b$ for these variables are given in Table
\ref{tvtable}.  

The local temperature $T$ is relevant to galaxy formation because gas which is
too hot does not satisfy the Jeans criterion and cannot cool efficiently, and
thus cannot collapse and form stars.  Since this criterion is explicitly
included in our conditions for the condensation of galaxy particles out of the
gas, it is a reasonable second variable to investigate. Figure~\ref{twovar.T}
shows a contour plot of galaxy density as a function of mass density and gas
temperature; all fields in this plot have been smoothed over 1 $h^{-1}$ Mpc
radius top hat spheres. First, note that local gas temperature and local dark
matter density are clearly not independent variables; the upper left triangle
in the figure is blank because there are no volume elements with high local
density and low temperature on these scales. Nevertheless, one can see that
galaxy density declines as the temperature increases at constant $\delta$.
Figure~\ref{tdep.T} shows $\delta_g$ as a function of $T$ for $\delta = 20$,
corresponding to a horizontal slice of Figure~\ref{twovar.T}.  The strength of
the temperature dependence is evident. From Table \ref{tvtable}, we note that
the inclusion of this extra variable reduces the variance by over a factor of
two, indicating that variations in local temperature can account for a large
portion of the stochasticity in the biasing relations of Figure \ref{alldd}. In
Figure \ref{tdep.T} we also plot the dependence on temperature of the galaxy
density in each age quartile, as the dashed curves.  As one would expect, the
density of the youngest galaxies is the most temperature-dependent; that is, in
the hottest regions, there are no new galaxies forming, although plenty of old
galaxies exist there, which formed there in the past or have since fallen
in. We have performed the same analysis using a simulation 50 $h^{-1}$ Mpc on a
side, with twice the resolution of our 100 $h^{-1}$ Mpc simulation. The
dependence of galaxy density on mass density and temperature is identical,
except for $\delta < 5$, where galaxies are produced more efficiently in the
higher resolution simulation. This difference has little effect on the
quantities $b$ and $r$, although the void probability function is affected, in
the sense that voids are less likely in the higher-resolution simulation.

A fairly good fit to the dependence of galaxy density on mass density and
temperature on small scales is:
\begin{equation} 
\label{paraboloid}
\frac{\rho_g }{\avg{\rho_g}}
= L \left(\frac{\rho}{\avg{\rho}} \right)^M \left(1+\frac{T}{40,000
\mathrm{~K}}\right)^N
\mathrm{.} 
\end{equation} 
We show such a fit in Figure \ref{tdep.T.3}. As labeled, each pair of
solid and dotted lines corresponds to a different value of
$\delta$. The solid lines are the results from Figure \ref{twovar.T};
the dotted lines are the fit of Equation (\ref{paraboloid}), using
$L=1.23$, $M=1.9$, and $N=-0.66$. For $N<0$, the factor involving $T$
takes into account the fact that relatively fewer galaxies have formed
in hotter regions; the form assumed here reflects the approximate
power law dependence in Figures \ref{tdep.T} and \ref{tdep.T.3}. This
effect is not important once the gas is as cold as $40,000$ K, and
thus we construct the temperature factor to have little effect in that
regime. 

We would like to be able to apply such a fit to $N$-body
simulations. Doing so would allow us to explore changes of
cosmological parameters more easily than the expensive hydrodynamical
simulations allow. Although we cannot follow gas temperature in purely
collisionless simulations, we can calculate the related quantity
$\avg{v^2}$, the local dark matter velocity dispersion. The dependence
on $\avg{v^2}$ should be similar to that on local gas
temperature. After all, in virialized regions the velocity dispersion
of dark matter particles is close to that of individual atoms. Indeed,
from Table \ref{tvtable} it is clear that $\sigma_{b,2}/\sigma_b$ is
nearly the same for the velocity dispersion as it is for the
temperature. Note, of course, that taking into account the dependence
of galaxy density on {\it both} $\avg{v^2}$ and $T$ would not improve
on using either one separately, as they are essentially equivalent
variables.

Next, we consider the oblateness and prolateness of the density field
in spheres of $1$ $h^{-1}$ Mpc, calculated from the eigenvalues of the
local inertia tensor $\lambda_1$, $\lambda_2$, $\lambda_3$; the
prolateness is $2\lambda_1/(\lambda_2+\lambda_3)-1$ and the oblateness
is $(\lambda_1+\lambda_2)/2\lambda_3-1$. As one can see from Table
\ref{tvtable}, these quantities are not important to the galaxy
density.  Finally, we consider the shear in the velocity field,
defined by:
\begin{eqnarray}
\Sigma^2 &=& \sum_{ij} (\Sigma_{ij})^2 \mathrm{,~where}\cr
\Sigma_{ij} &=& \frac{1}{2} (v_{i,j}+v_{j,i}) - \frac{1}{3} v_{k,k}
\delta_{ij} \mathrm{.}
\end{eqnarray}
Clearly the shear will correlate well with the dark matter density, as the dark
matter in the higher density regions will have higher velocities. To account
for this, we consider the dimensionless quantity
$\Sigma^2/H_0^2(1+\delta)$. Another way to justify this scaling is to think of
the $\Sigma$ as a characteristic shearing rate, to be multiplied by the
dynamical time, which scales as $(1+\delta)^{-1/2}$.  From Table \ref{tvtable},
the galaxy density seems largely independent of the shear in the velocity
field.  Thus temperature, or equivalently velocity dispersion, seems to be the
second parameter which best minimizes the stochasticity of the galaxy-mass
relationship. There may be additional variables important to the process of
galaxy formation, but this set of variables is fairly broad and includes the
most likely candidates.

\section{Explaining Scale Dependence}
\label{toy}

The last section revealed that local temperature is an important
parameter for determining the local galaxy density. Is it possible
that the temperature dependence can affect the scale dependence of
$b(R)$? That is, as we pointed out in the last section, does the
temperature have statistical properties similar to those of the
scatter around the conditional mean $\avg{\delta_g|\delta}$?  We can
repeat the exercise of Section~\ref{onevar} and calculate a fake
galaxy density field, this time using the two-variable mean
$\avg{\delta_g|\delta, T}$ of Figure~\ref{twovar.T}. We then calculate
$b(R)$ as before and plot it as the short dashed line in each panel of
Figure~\ref{bth}. The temperature dependence indeed beautifully
accounts for the variation of $b(R)$ with scale, for all the galaxies
as well as for each quartile. The fit to the density distribution
given by Equation (\ref{paraboloid}) produces nearly identical
results; it is shown as the long dashed line in the upper panel.

What are the physical properties of the temperature field that cause
this to happen?  Essentially, it is that local temperature reflects
the gravitational potential and thus contains information about the
large scale density field. The temperature fluctuations $\delta_T \equiv
T/\avg{T} - 1$ can be expressed in a simple way by considering some
limits (\cite{spergel98ap}). First, in the nonlinear regions, gas is
virialized and its temperature must scale as the local potential. From
Poisson's equation, one knows that $\tilde\phi(\vv{k}) \propto k^{-2}
\tilde\delta(\vv{k})$. Thus, on small scales, it must be that
$\tilde\delta_T(\vv{k})\propto k^{-2} \tilde\delta(\vv{k})$ as
well. Second, on linear scales one can assume that the temperature
fluctuations are dominated by the number density fluctuations of
virialized haloes, since gas in those areas is much hotter than that
in the empty regions between halos. Thus, on large scales
$\tilde\delta_T(\vv{k})\propto\tilde\delta(\vv{k})$. These two limits
may be combined:
\begin{equation}
\label{deltaT}
\tilde\delta_T(\vv{k}) \propto
\frac{\tilde\delta(\vv{k})}{1+k^2r_{nl}^2/(2\pi)^2} \mathrm{,}
\end{equation}
where $r_{nl}$ is the transition scale between the linear and
nonlinear regimes. Consequently, we expect the cross spectrum of
temperature and mass to be:
\begin{equation}
\label{ptt}
P_{Tm}({k}) \propto
\frac{P_{mm}({k})}{1+k^2r_{nl}^2/(2\pi)^2} \mathrm{.}
\end{equation}
In Figure \ref{pstemp}, we compare this simple model with the
simulations, finding that it is a good fit for the choice $r_{nl} =
16$ $h^{-1}$ Mpc. This value for $r_{nl}$ agrees approximately with
the scale on which nonlinear effects should become important.  Thus,
the temperature power spectrum peaks at large scales; furthermore, at
those scales the temperature fluctuations are directly
related to the mass density. That means that the
largest contribution to the local temperature actually comes from
large wavelength fluctuations, which follow the large wavelength
fluctuations in mass density. The local gas
temperature is therefore a direct indicator of the large scale density
field. Thus, accounting for the temperature dependence automatically
accounts for the dependence on large scale density, and consequently
the dependence of bias on scale.

A simple model for the relation between galaxies, mass, and
temperature reveals more explicitly how scale dependence and
stochasticity enter the relation between galaxies and mass. Consider
the fit given in Equation (\ref{paraboloid}). If one assumes that
$\delta_g\ll 1$, $\delta\ll 1$, and $\delta_T\ll 1$, this relation
becomes:
\begin{equation}
\label{lin2}
\delta_g = M'\delta + N'\delta_T \mathrm{,}
\end{equation}
and one recovers the deterministic linear bias model if $N'=0$ and
$M'=b$.  This model is obviously highly unrealistic, especially at
small scales. However, its simplicity will allow us to understand
better how $b(R)$ becomes scale-dependent.

We can perform a linear regression on $\delta$ and $\delta_T$ to
determine $M'$ and $N'$. The results as a function of top hat
smoothing scale are shown in Figure~\ref{bdbt}. Notice that $M'$ and
$N'$ are approximately constant with respect to scale. This invariance
indicates that the scale dependence is well-accounted for by the
temperature dependence.  To examine this claim, let us assume that
Equation (\ref{lin2}) holds at some small scale $R_0$.  Now, if we
smooth over a larger scale $R$, we find that:
\begin{equation}
b^2(R) \equiv \frac{\sigma_g^2(R)}{\sigma^2(R)} = M'^2 + N'^2
\frac{\sigma_T^2(R)}{\sigma^2(R)} +
2M'N'\frac{\avg{\delta\delta_T}_R}{\sigma^2(R)} \mathrm{.}
\end{equation}
Note that $\avg{\delta\delta_T}_R/\sigma^2(R)$ and
$\sigma_T^2(R)/\sigma^2(R)$ will depend on $R$,
because Equation (\ref{deltaT}) shows that $P_{Tm}(k)/P_{mm}(k)$ and
$P_{TT}(k)/P_{mm}(k)$ depend on $k$.  Thus,
$b(R)\equiv\sigma_g(R)/\sigma(R)$ will also depend on $R$. We can test
this possibility directly by applying the deterministic two-variable
linear model to this simulation and calculating $b(R)$. For this test
we use the values $M'=2.4$ and $N'=-0.4$, which are appropriate at 1
$h^{-1}$ Mpc; the resulting curve is the dot-dashed line in the upper
panel of Figure
\ref{bth}. At small scales, the linear approximation is (as expected)
poor but of the right order; on large scales, it reproduces the value
of $b(R)$ fairly well.

Thus, even this simple linear model reproduces the scale dependence.
We do not claim that this model is a particularly {\it good} one for
describing the galaxy density; the fit of Equation (\ref{paraboloid})
is much better. Instead, the linear model is merely a toy which
illustrates the following point: given a dependence of galaxy density
on local temperature, it is inevitable that $b(R)$ is a function of
scale on scales smaller than about $r_{nl}$, simply because of the
relationship between gas temperature and mass density.

\section{Discussion}

\label{discussion}

Consider two regions of the universe, both of which have the same
local mass overdensity; however, with respect to the large scale
density field, one is in an overdensity, the other in a void. The
gas in these two regions will evolve similarly, forming galaxies with
about equal efficiency, until the large scale density field becomes
nonlinear. At this point, the ambient gas around the first region, in
the large scale overdensity, will become too hot to accrete any longer,
and galaxy formation will cease. Meanwhile, the gas around the second
region, in the void, will remain cool enough to accrete onto old
galaxies and continue to form new ones. This picture explains part of
the scatter in the local relation between $\delta$ and $\delta_g$; it
also explains why this scatter is such that in hot regions $\epsilon
\equiv \delta_g - \avg{\delta_g|\delta}<0$, and in cold regions
$\epsilon>0$. In turn, the correlation of temperature with large scale
density explains why $\epsilon$ also correlates with large scale
density, causing scale dependent bias. In addition, it indicates
qualitatively why spirals, which are caused by late-time accretion of
gas, are relatively more abundant in the ``field'' and relatively
underabundant in the rich clusters.

In this scenario, scale dependent bias follows from only one rather
robust assumption: that the ambient gas temperature affects the
efficiency of galaxy formation. Since gas must be able to cool to form
galaxies, this assumption is well-motivated
theoretically. Furthermore, observations of the star-formation rate as
a function of local density indicate that, indeed, star formation is
reduced in the hot cluster environments, even at a fixed morphological
type (\cite{young96ap}; \cite{hashimoto97ap}; \cite{balogh98ap}). Once
one concedes that local temperature is an indicator of the 
efficiency of galaxy formation, the arguments in Section
\ref{toy} lead one to directly conclude that $b(R)$ should depend on
scale, at least in the quasilinear and nonlinear regimes, where
$P_{Tm}(k)$ and $P_{TT}(k)$ have different shapes than $P_{mm}(k)$. In
the simulations studied here, the effects are particularly strong.

\begin{figure}
\figurenum{\fignum b}
\epsscale{1.0}
\plotone{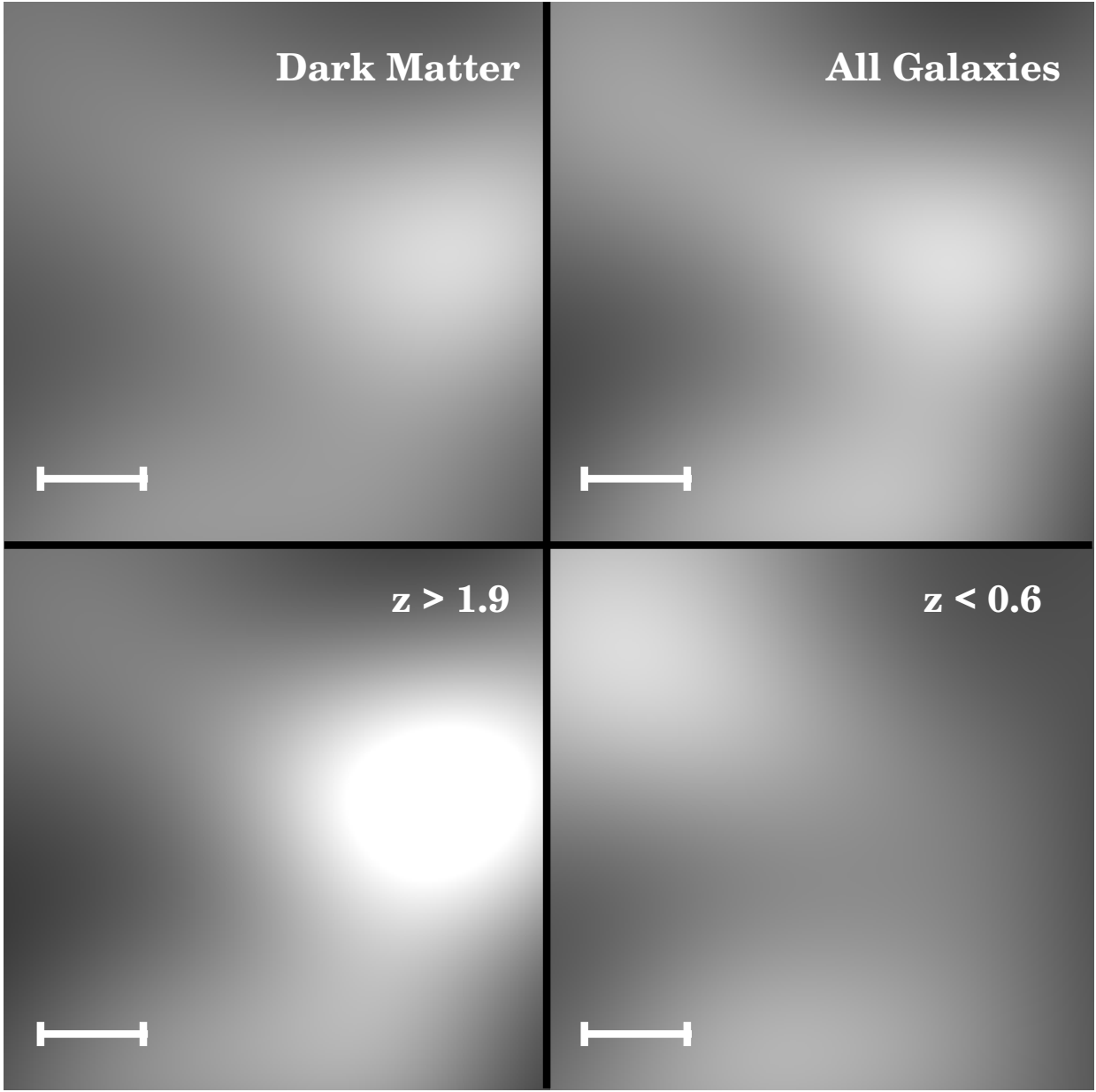}
\epsscale{1.0}
\caption{\label{quad.ls} Same as Figure \ref{quad.ss}, now using 10
$h^{-1}$ Mpc Gaussian smoothing. It is apparent now that on these
scales the oldest galaxies are highly biased and that the young
galaxies are poorly correlated with the mass. They appear to be
overdense along the filaments but underdense in the clusters. The
young galaxies appear quite different than one would guess based on
Figure \ref{quad.ss}, but note that that figure has a logarithmic
stretch and that structure outside the two-dimensional slice
contributes significantly to the smoothed field in this
figure.}
\end{figure}

\begin{figure}
\figurenum{\fignum}
\epsscale{0.85}
\plotone{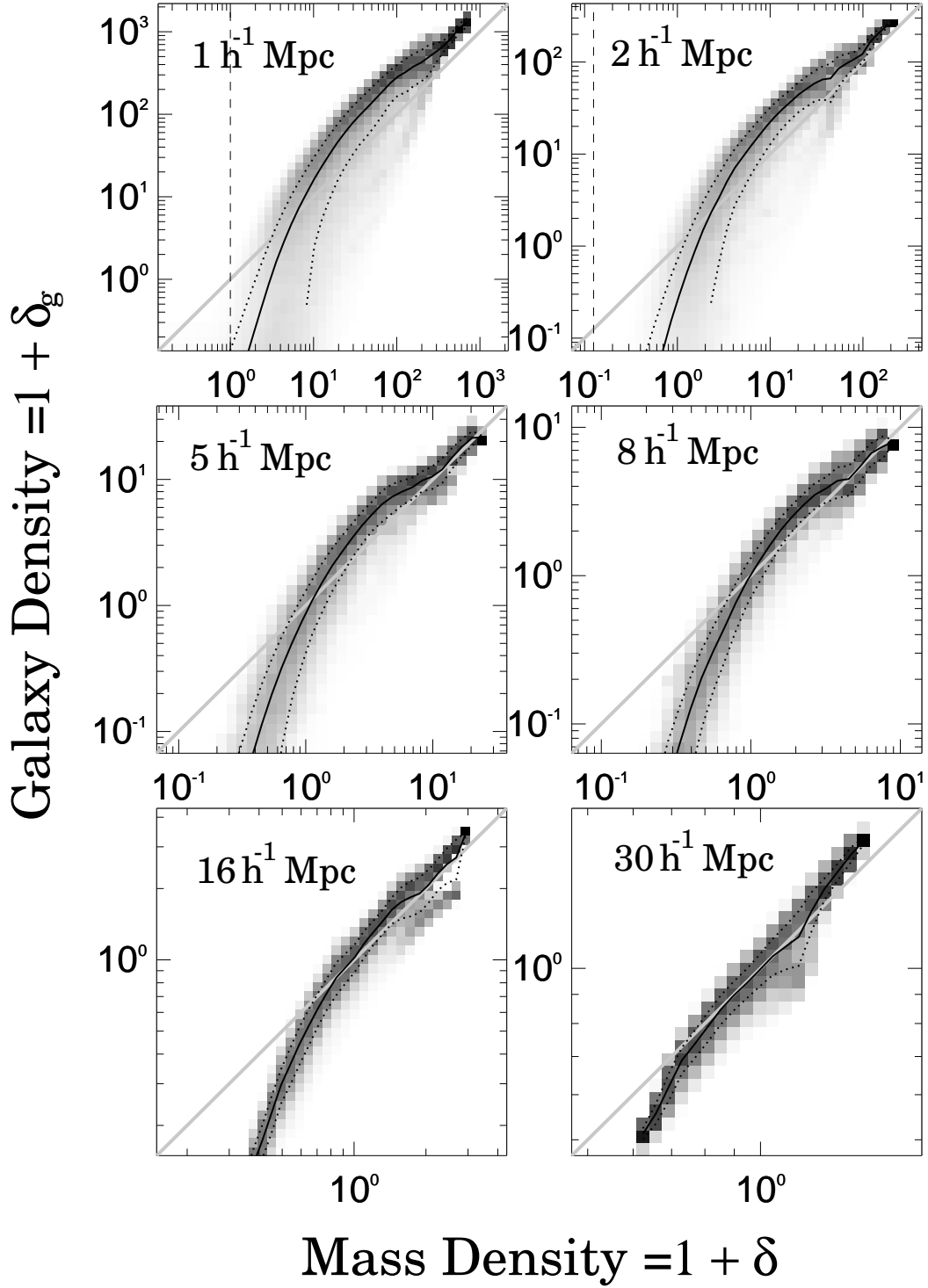}
\epsscale{1.0}
\caption{\label{alldd} Galaxy mass density as a function of mass
density for a variety of top hat smoothing radii. The shading is a
logarithmic stretch of the conditional probability $P(1+\delta_g |
1+\delta)$; thus, each column is normalized separately. The solid
black lines indicate $\avg{1+\delta_g|1+\delta}$; the dotted black
lines indicate the 1$\sigma$ deviation from the mean.}
\end{figure}

\begin{figure}
\figurenum{\fignum a}
\plotone{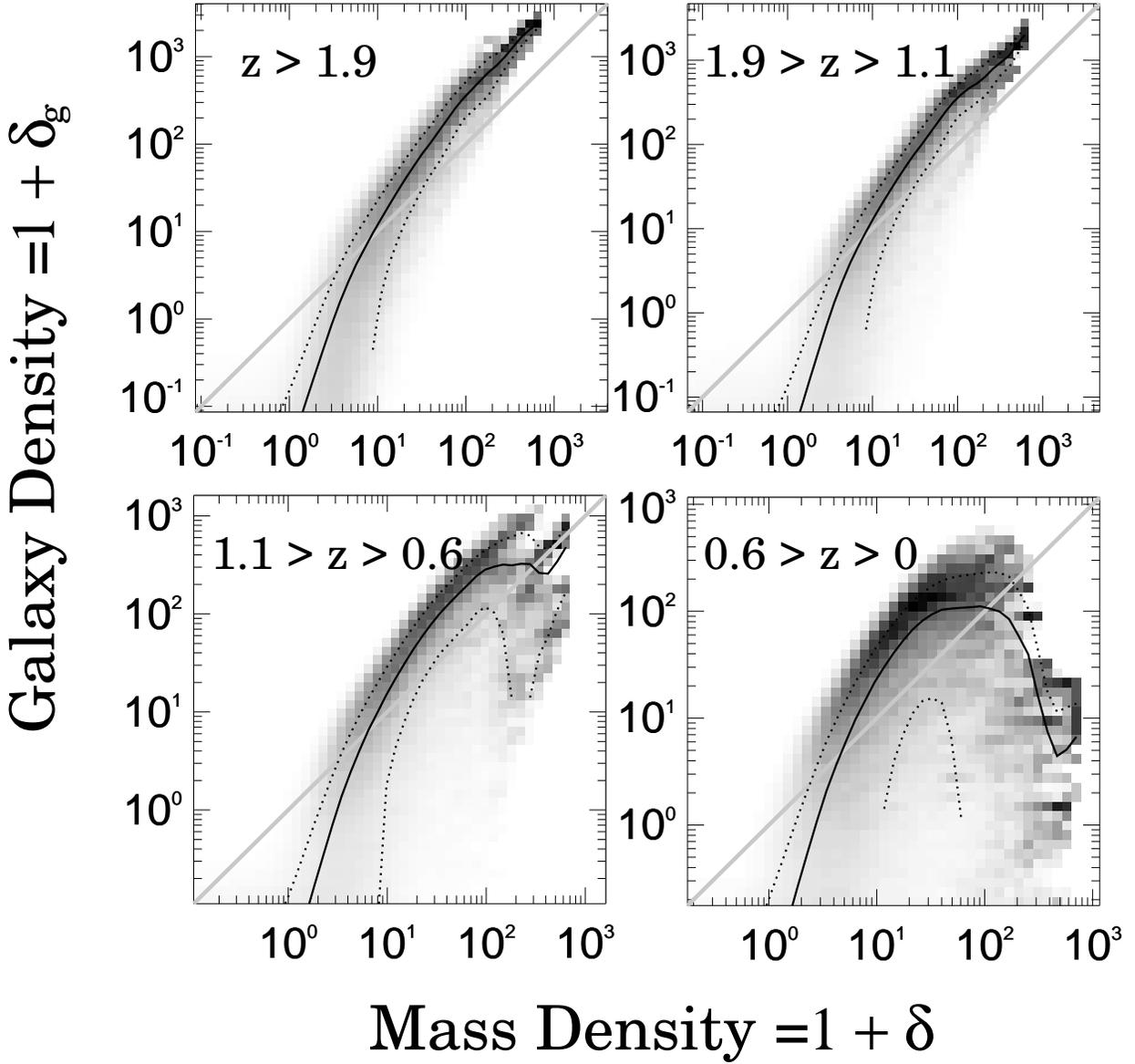}
\caption{\label{agedda} Galaxy mass density as a function of dark
matter density for each age quartile. Smoothing filter is a 1 $h^{-1}$
Mpc radius top hat sphere. In each panel we list the range of
formation redshifts included. The plot is of the same form as
Figure~\ref{alldd}.}
\end{figure}

\begin{figure}
\figurenum{\fignum b}
\plotone{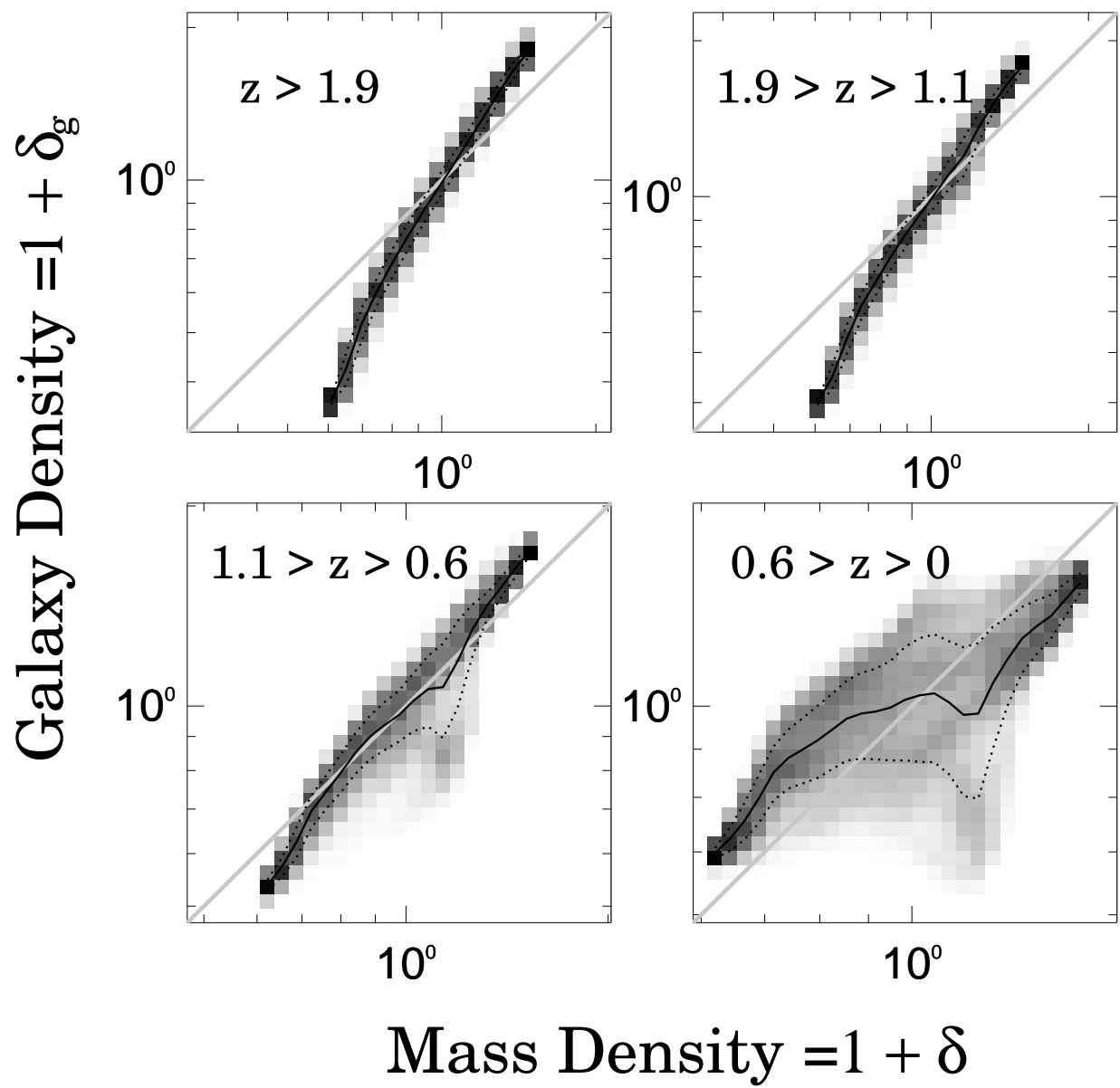}
\caption{\label{ageddb} 
Same as Figure \ref{agedda}, with a 30
$h^{-1}$ Mpc radius top hat smoothing filter. }
\end{figure}

\begin{figure}
\figurenum{\fignum}
\plotone{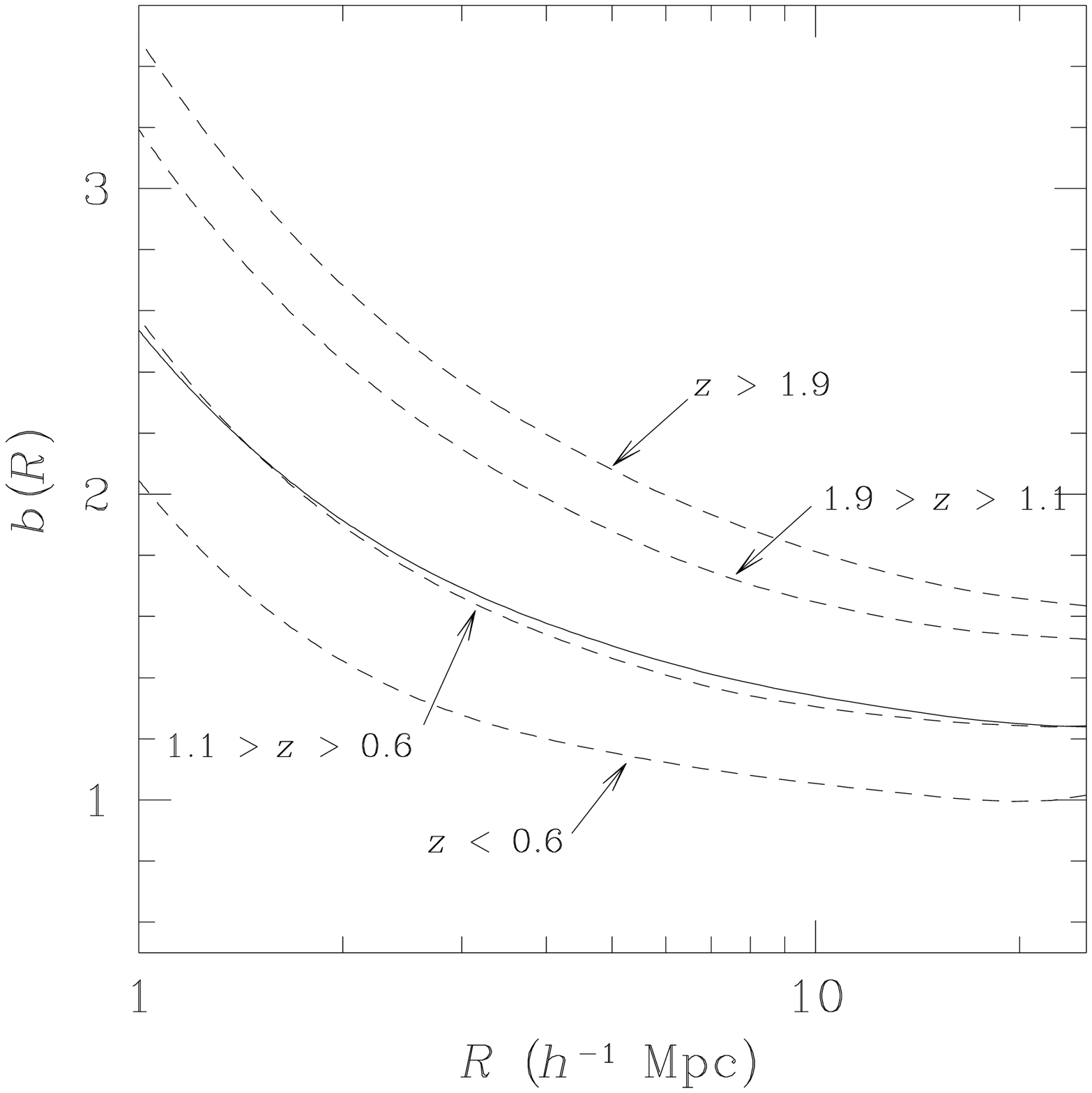}
\caption{\label{b} The bias $b(R)\equiv\sigma_g(R)/\sigma(R)$. The solid
line indicates all the galaxies. The dashed lines indicate the age
quartiles, with range of formation redshifts listed. Note the strong
scale-dependence, and the fact that old galaxies are more biased than
young galaxies.}
\end{figure}

\begin{figure}
\figurenum{\fignum}
\plotone{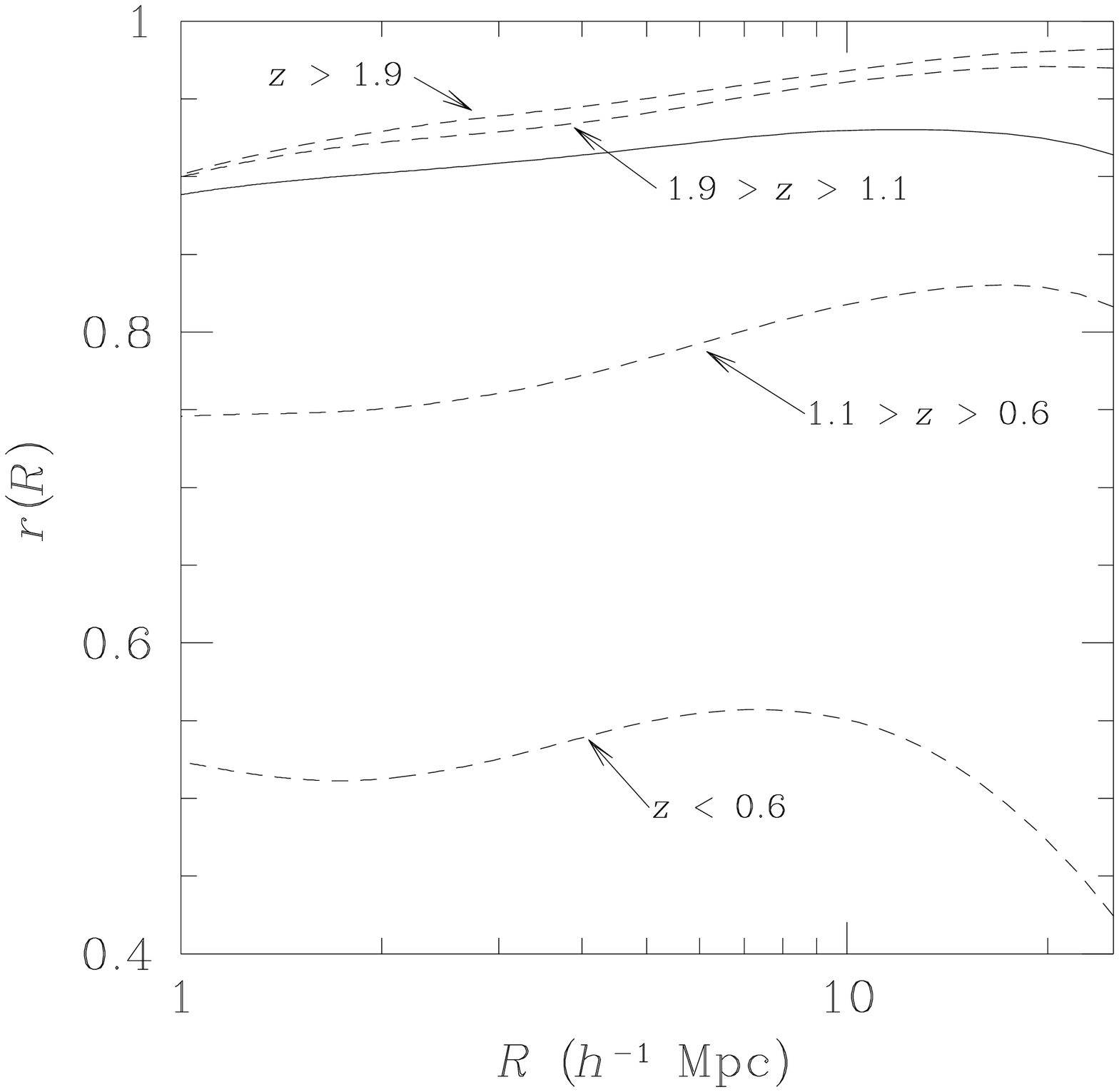}
\caption{\label{r} The correlation coefficient
$r(R)\equiv\avg{\delta_g\delta}/\sigma_g(R)\sigma(R)$ as defined in the
text. The lines have the same meanings as in Figure \ref{b}.
The youngest galaxies are poorly correlated with the underlying mass
distribution at all scales. For the oldest galaxies, the decline at
small scales of the correlation coefficient is probably mainly due to
the nonlinearity of $\avg{\delta_g|\delta}$.}
\end{figure}

\begin{figure}
\figurenum{\fignum}
\plotone{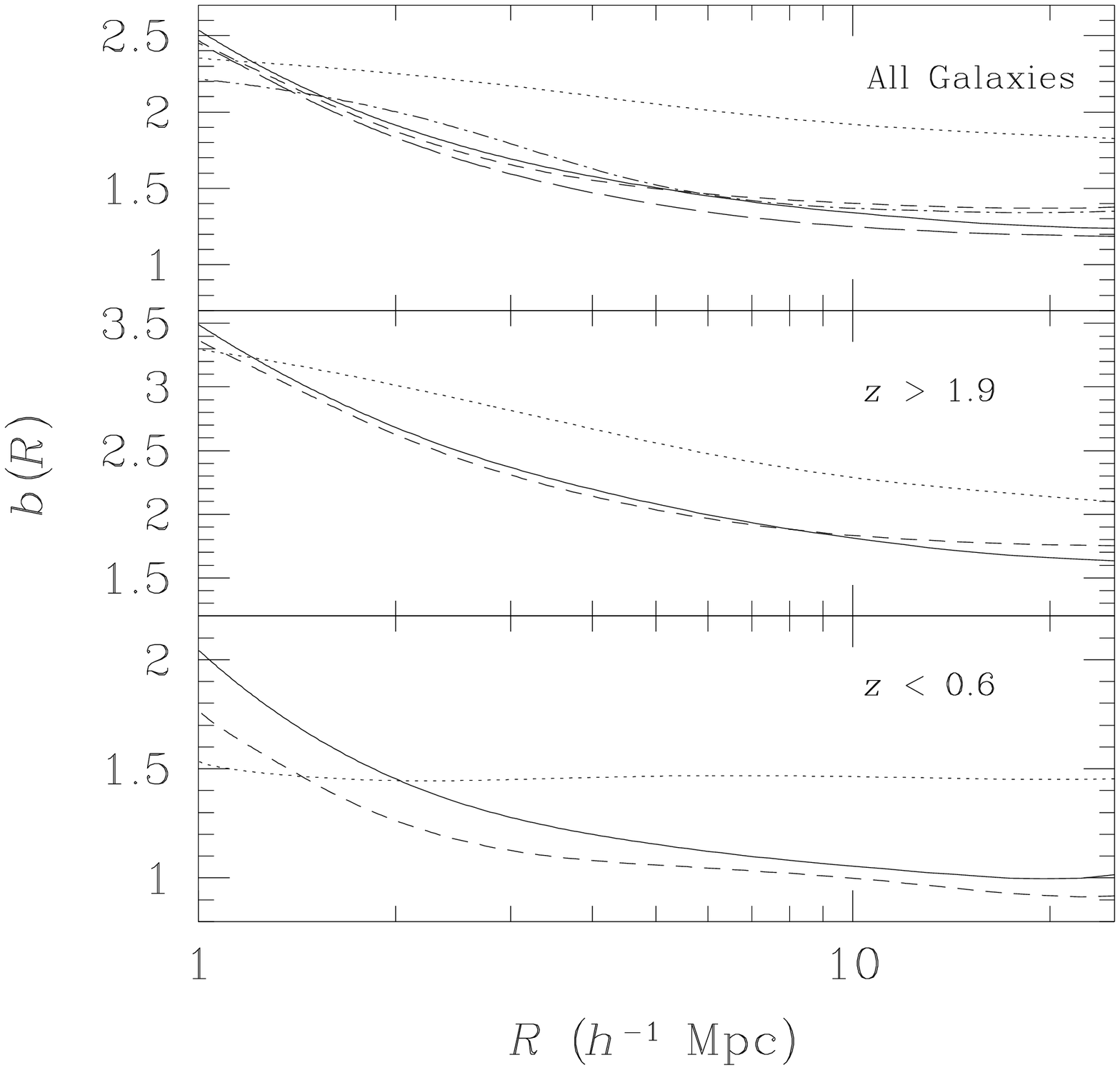}
\caption{\label{bth} The bias $b(R)$ for various bias models compared
with $b(R)$ for the galaxy particles in the simulations. As labeled,
the top panel refers to all the galaxies, the middle panel to the
oldest quartile of galaxies, and the bottom panel to the youngest
quartile of galaxies.  The solid lines are for the actual galaxy
particles. The dotted lines are for a galaxy field defined by
$\avg{\delta_g|\delta}$, with a 1 $h^{-1}$ Mpc tophat smoothing. The
short dashed lines are for a galaxy field 
defined by $\avg{\delta_g|\delta,T}$. The long dashed lines are for a
galaxy field defined by the fit to $\avg{\delta_g|\delta,T}$ given in
Equation \ref{paraboloid}. The dot-dashed line is for a
galaxy field defined by the toy model in Equation (\ref{lin2}), with
$M'=2.4$ and $N'=-0.4$. All the two-variable models reproduce the
scale-dependence much better than does the single-variable model.}
\end{figure}

\begin{figure}
\figurenum{\fignum}
\plotone{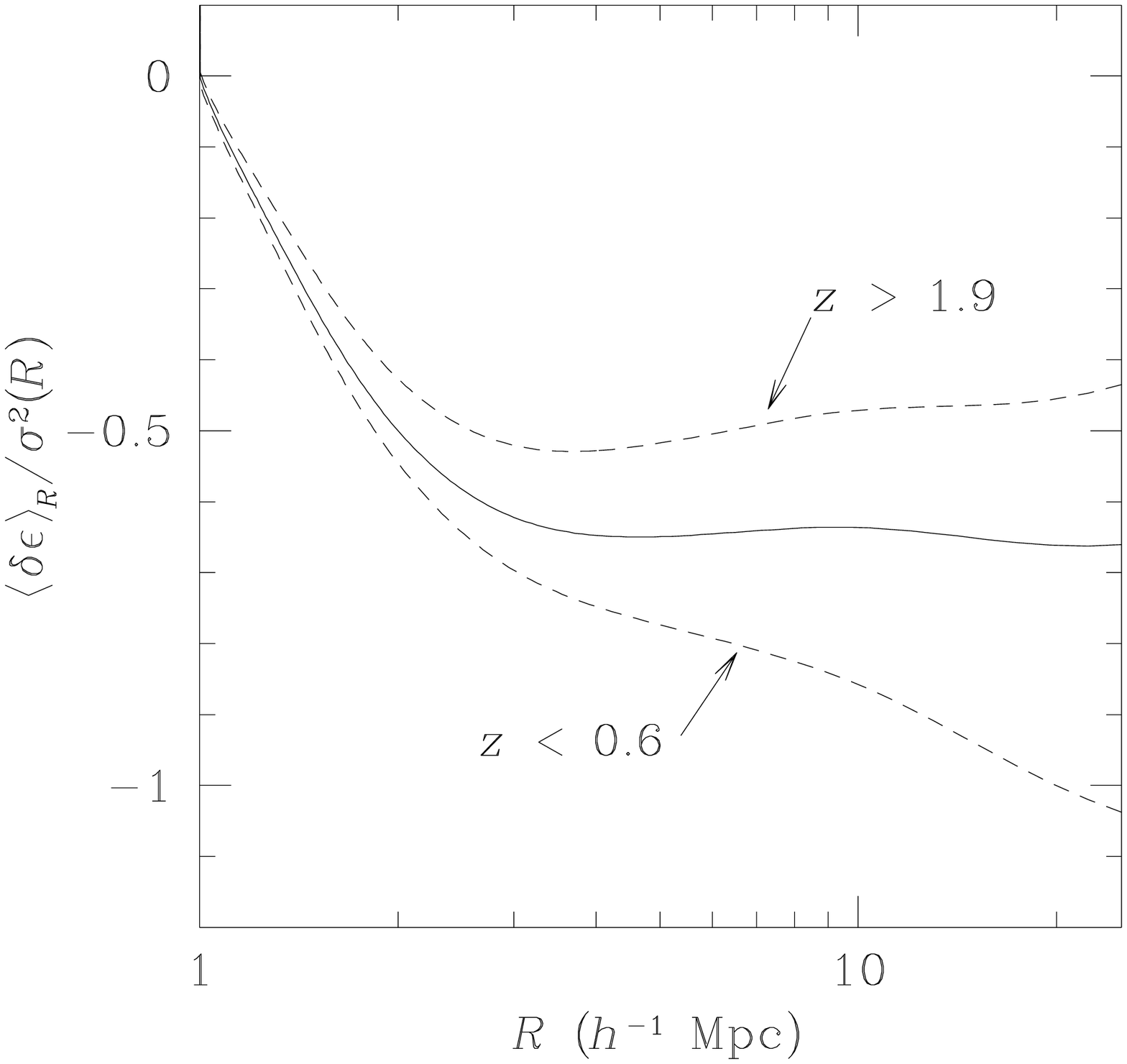}
\caption{\label{epsdel} The correlations between the density field
$\delta$ and the residuals $\epsilon\equiv\delta_g-\avg{\delta_g|\delta}$
defined on 1 $h^{-1}$ Mpc scales, as a function of scale. The solid
curve is for all galaxies, and the dashed are for the oldest and
youngest quartiles, as labeled.  As explained in the text, the
correlation of residuals with the large scale field is an indication
that $b(R)$ will be a function of scale. Also note that the dependence
is stronger for the young galaxies than for the old.}
\end{figure}

\begin{figure}
\figurenum{\fignum}
\plotone{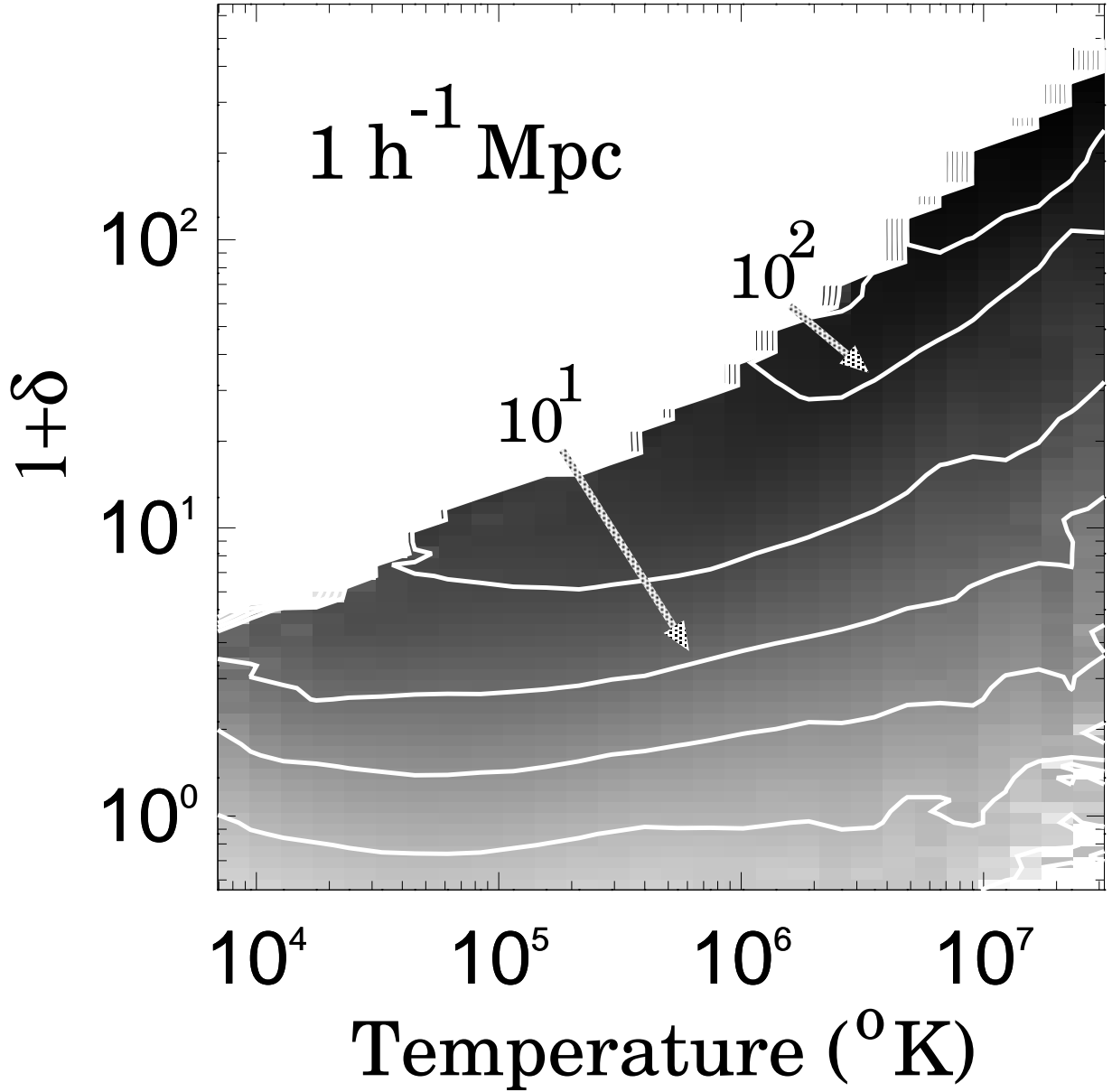}
\caption{\label{twovar.T} Dependence of galaxy density $1+\delta_g$ on
dark matter density $1+\delta$ and temperature $T$, evaluated at 1
$h^{-1}$ Mpc smoothing. The greyscale is a logarithmic stretch of
$1+\delta_g$; the contours are in even logarithmic intervals. }
\end{figure}

\begin{figure}
\figurenum{\fignum}
\plotone{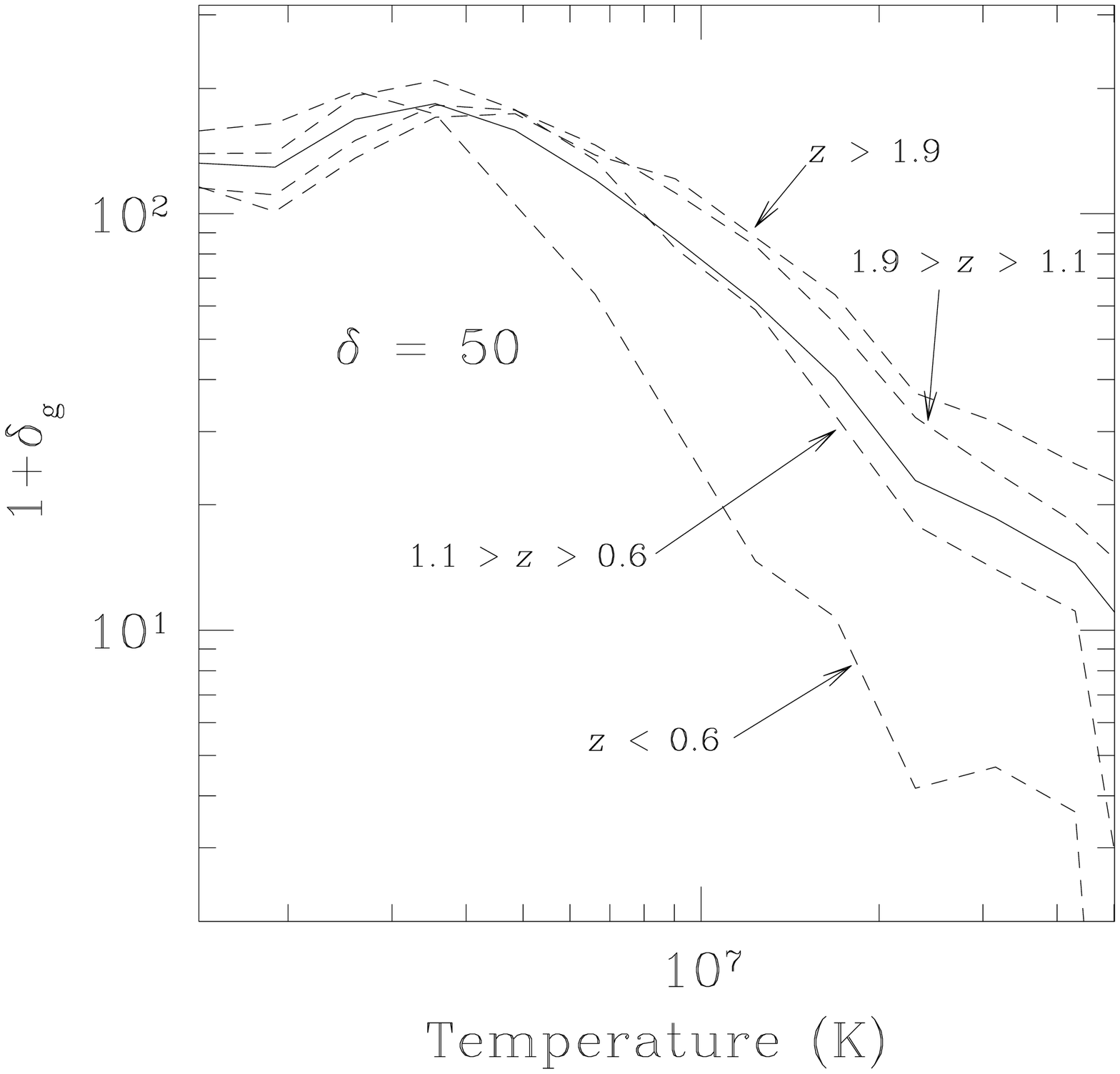}
\caption{\label{tdep.T} The dependence of $\rho_g/\avg{\rho_g}\equiv
1+\delta_g$ on $T$ at fixed $\delta$ for all the galaxies (solid line)
and each quartile (dashed lines, as labeled). Note that
$(1+\delta_g)$ varies over an order of magnitude in this
temperature range. Also, note that the young galaxies have the strongest
dependence on temperature and the oldest galaxies have the weakest.}
\end{figure}

\begin{figure}
\figurenum{\fignum}
\plotone{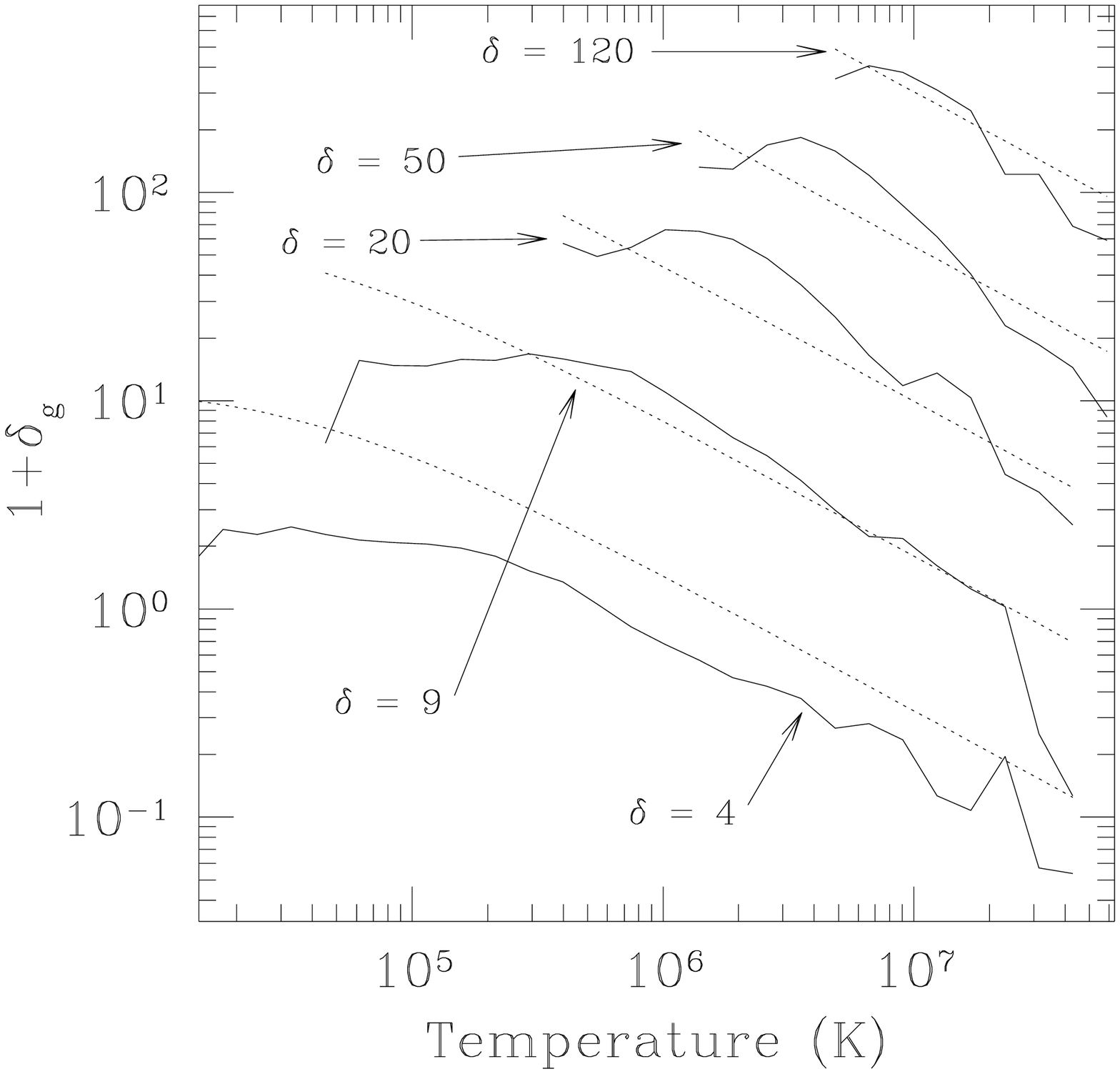}
\caption{\label{tdep.T.3} The dependence of
$\rho_g/\avg{\rho_g}\equiv 1+\delta_g$ on $T$ at a series of fixed $\delta$
for all the galaxies. The solid lines are the actual dependence of the
galaxies in the simulations found in Figure \ref{twovar.T}. The dotted
lines are from the fit in Equation (\ref{paraboloid}), with the
parameters $L=1.23$, $M=1.9$, and $N=-0.66$.}
\end{figure}

\begin{figure}
\figurenum{\fignum}
\plotone{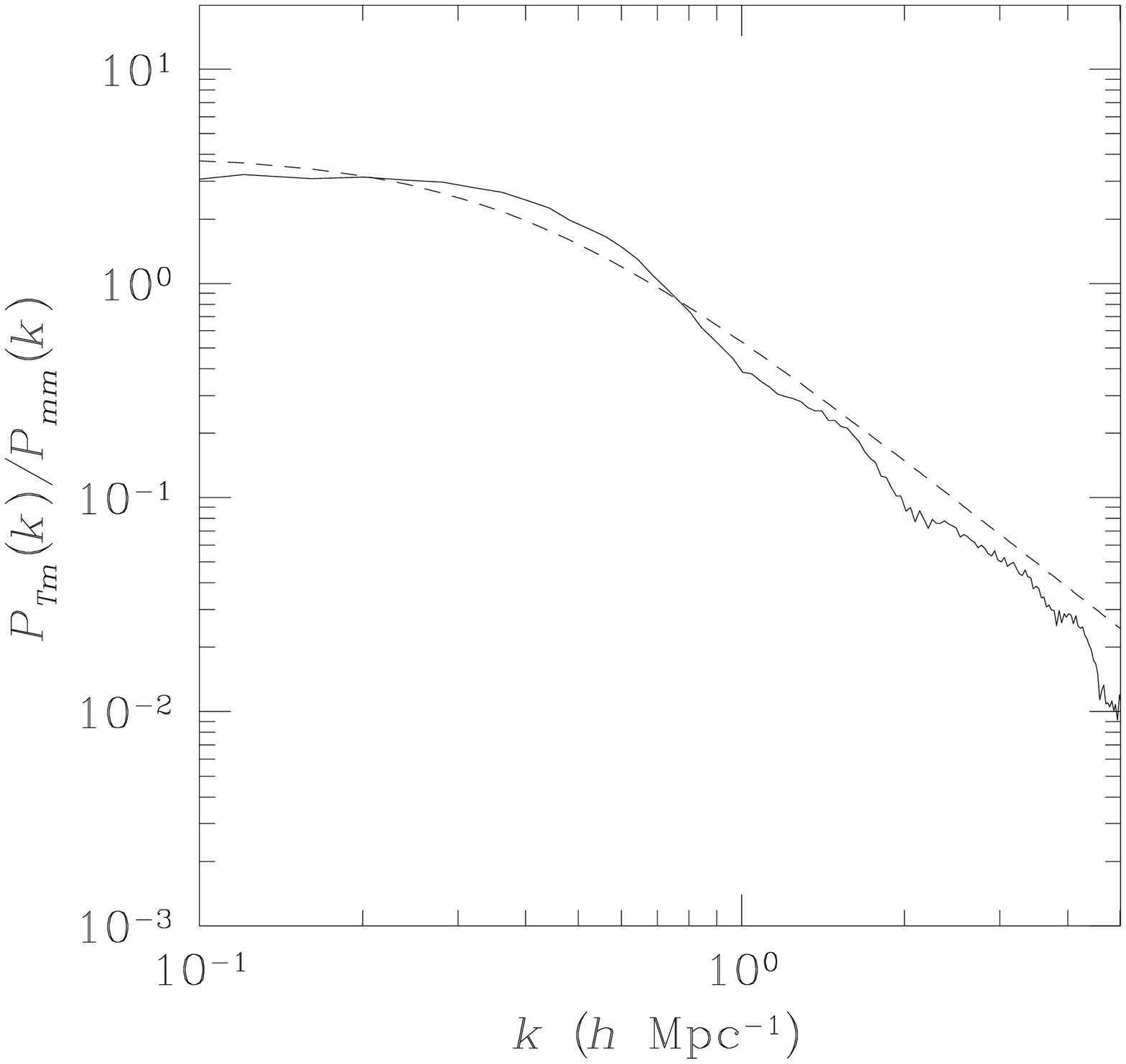}
\caption{\label{pstemp} Comparison of the simple model for the
temperature-density cross spectrum given in Equation \ref{ptt} with
the cross spectrum measured in the simulations. The solid line is from
the simulations and the dashed line is the model, for $r_{nl} = 16$
$h^{-1}$ Mpc.}
\end{figure}

\begin{figure}
\figurenum{\fignum}
\plotone{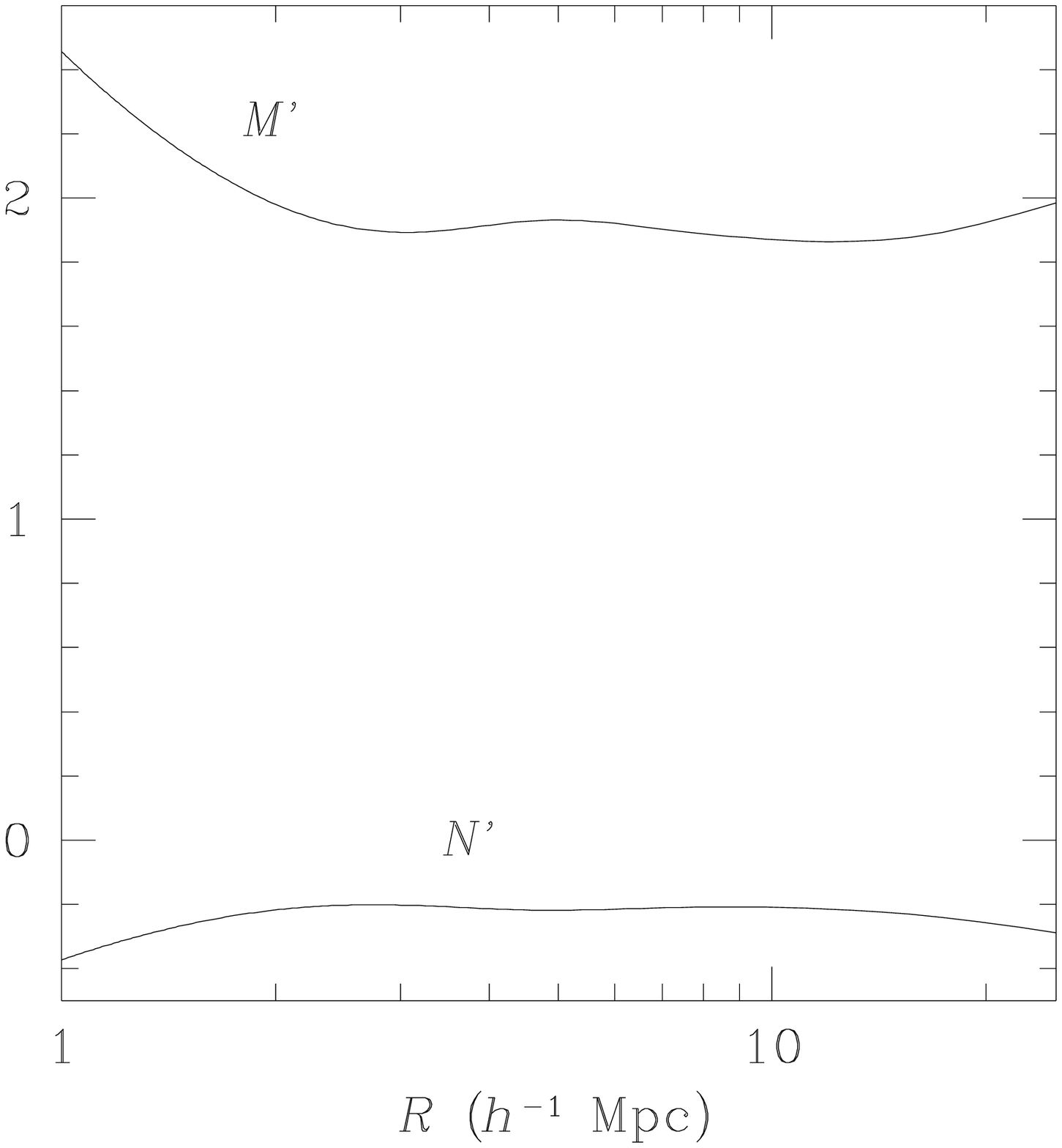}
\caption{\label{bdbt} The linear parameters $M'$ and $N'$ of Equation
(\ref{lin2}), as a function of scale. We calculated these using a
linear regression on $\delta$ and $\delta_T$. Note the scale
independence of this model, as compared to the scale dependence in the
single-variable model parameter $b(R)$, shown in Figure \ref{b}.}
\end{figure}

Scale dependence in the relation between galaxies and mass can affect
the interpretation of future redshift and peculiar velocity
surveys. The most obvious example is that on small scales, the shape
of the galaxy power spectrum will differ from that of the mass.
Furthermore, as pointed out by \cite{dekel98a}, comparison of the
observed galaxy density field to that inferred from observed peculiar
velocities (\cite{dekel94ap}; \cite{sigad98ap}) effectively perform a
regression of $\delta$ on $\delta_g$ and thus measure the quantity
$\beta=(r/b)f(\Omega)$, which we show here depends both on scale and
on the chosen galaxy sample. Thus, the current analyses, usually
performed in the quasilinear regime and using \iras\
galaxies\footnote{\iras\ galaxies are typically young, although they
are probably not as young as our youngest quartile.  In particular,
\iras\ galaxies do not show an underdensity in rich clusters that the
youngest galaxies in the simulations do (Fig.~\ref{quad.ls}), although
they are less overdense than are optically selected galaxies (Strauss {\it
et al.}~1992a).} to define the density field, may be sensitive to
these effects.  For instance, one's estimate of $r/b$ will generally
increase with scale, by about 20\% between 5 $h^{-1}$ Mpc and $30$
$h^{-1}$ Mpc.  Amusingly, although $r$ and $b$ vary by a factor of two
between the youngest and oldest galaxies on large scales, the
dependence of $r/b$ on galaxy age remains quite small; $r/b \sim
0.6$--$0.7$ for all four quartiles. In the real world, the regression
of $\delta$ on $\delta_g$ will most likely not be this constant among
the morphological types. The decrease of $\beta$ with scale could
contribute to the differences between the results of \cite{sigad98a}
and of analyses carried out with smaller smoothing scale, such as
VELMOD (\cite{willick98ap}); however, there are stark differences
between these methods, and a more careful analysis is thus necessary.
Indeed, we are interested in exploring the effects of nonlinear
stochastic bias on a variety of large-scale structure statistics
inferred from redshift surveys, including redshift-space distortions
(\cite{dekel98ap}) and the pairwise velocity dispersion as a function
of local density (\cite{strauss98ap}).  A full treatment will require
identifying individual galaxies from the galaxy particles in the
simulations, which will require tackling the overmerging problem in
the densest regions of the simulations. 

We can address some of these issues without having to run expensive
hydrodynamic simulations for a range of cosmological models. First, in order to
examine in detail the effect of the relation between galaxy density, mass
density, and temperature on all of these statistics, we plan to carry out
$N$-body simulations of larger dynamic range than is possible with the current
hydrodynamical simulations. To characterize the galaxy distribution in these
simulations, we will apply the model of Equation (\ref{paraboloid}), using
$\avg{v^2}$ as a proxy for temperature. Then one can explore the effect this
type of bias can have on the various statistics discussed in the last
paragraph. A second approach is to analyze observations, allowing for scale-
and temperature-dependent bias of the character described here. In this vein,
one could investigate the differences between galaxy types and see how they
compare in detail with the differences we find between our age quartiles. Using
the local galaxy velocity dispersion as a proxy for temperature, we could
even directly investigate the dependence of the density of difference
morphological types on temperature.

Of course, that temperature is the important \emph{causal} variable in these
simulations is based only on a {\it post hoc} (though physically plausible)
argument in this paper. Any variable that probes the large-scale density field
would serve just as well to reduce $\sigma_{b,2}/\sigma_b$ and explain the
scale dependence of $b(R)$. However, as discussed above, temperature is a
well-motivated quantity.  Controlled tests of the effect of ambient temperature
on star formation in these simulations might help clarify the matter. Another
approach is to look at various output times and examine under what conditions
in the simulations the galaxies actually form. We do so in a separate paper,
which examines the time dependence of galaxy formation and bias in these
simulations. That work makes it clear that the temperature dependence is a
result of the Jeans mass and cooling criteria that the code uses to decide
where galaxy particles condense out of the gas.

The alert reader will notice that some of our results contradict our
opening statements in Section \ref{motiv}. In particular, if one looks
at our results on large scales, it does happen to be true that the
distributions of old galaxies and of young galaxies both differ
considerably from the mass distribution, but in combination trace the
mass quite well and are almost unbiased.  We simply note here that the
same would not be true were we to look at the results of these
simulations at $z=0.5$, for instance. In addition, \cite{cen92b} found
that for a hot dark matter universe (HDM), $b$ was significantly
greater than unity on large scales, indicating that the level of bias
in these simulations depends somewhat on the chosen cosmology.  Thus
we ascribe little importance to this coincidence in the current
simulations.  

Note also that these results concern \emph{stellar mass density}, not
galaxy number density. Since the galaxies in the densest regions of
the simulations overmerge, to what degree the bias described here
affects the brightness, rather than the number, of galaxies is
unknown.  Only if the stellar mass function of galaxies is universal
will the bias found here translate directly into galaxy number density
bias. On the other hand, previous theoretical results hint that the
most massive halos form preferentially in large-scale overdensities
(\cite{mo96ap}).  Meanwhile, these massive halos may form fewer stars
per unit mass than less massive halos (\cite{white91ap};
\cite{katz92ap}; \cite{evrard94ap};
\cite{kauffmann98ap}). Observationally, \cite{bromley98a} find that
early-type galaxies are fainter in dense regions than elsewhere. The
temperature dependence found in our simulations would certainly
explain the last two effects, which do affect the brightnesses, but
not the numbers, of galaxies.

Several further notes of caution are in order concerning applying these
results in detail. First, the simulations have limited resolution and
do not probe physics on scales less than 200 $h^{-1}$ kpc. Thus, we
cannot follow individual halos in dense regions. As a check, we
performed the same analysis on a 50 $h^{-1}$ Mpc simulation of twice
the resolution, finding consistent results in the regime $\delta > 5$
for the dependence of galaxy density on mass density and temperature,
smoothed at 1 $h^{-1}$ Mpc. On the other hand, the 50 $h^{-1}$ Mpc
simulations produced more galaxies in the regions with $\delta <
5$. Second, by necessity all the complications of the interstellar
medium (ISM) and small scale dynamics are ignored.  Surely the
morphology of the ISM, its interaction with infalling gas, and its
reaction to high star-formation rates is important to the formation
and evolution of galaxies. Third, even to the extent that the
simulation is physically accurate, the dependence of the results on
cosmological parameters is unknown.  Finally, we have not experimented
here with varying the galaxy formation and feedback parameters,
although this has been done in previous work (\cite{cen92bp};
Gnedin 1996a,b). We plan future simulations of
higher resolution both to probe variations in cosmology, and to attempt
to identify individual galaxies even in the denser regions of the
simulation. In addition, we are working on methods of efficiently
exploring the effects of varying the galaxy formation parameters.

The general result of the simulations is that the relationship between
mass and galaxies is interestingly complicated. Given the precision
and volume of upcoming redshift surveys, it is possible that our
understanding of the mass distribution on small scales will be limited
by our ignorance of the properties of galaxy formation and of the
origin of different morphological types. Hopefully, cosmological
simulations such as this one can help us understand in what ways, and
possibly to what degree, the galaxy distribution can differ from that
of the mass.

\acknowledgments

This work was supported in part by the grants NAG5-2759, AST93-18185
and AST96-16901, and the Princeton University Research Board. MAS
acknowledges the additional support of the Alfred P.~Sloan Foundation,
and Research Corporation. We would like to thank David N.~Spergel for
useful discussions, and our editor, Ethan Vishniac, and an anonymous
referee for useful comments.

%
%

\newpage
\setcounter{thetabs}{1}

\begin{deluxetable}{ccc}
\tablecolumns{5}
\tablenum{\tabnum}
\tablecaption{\label{quartiles}
Age and redshift ranges of galaxy particle quartiles of equal total mass.}
\tablehead{Quartile & Age Range (Gyrs) & Redshift Range}
\startdata
Oldest           & 9.6 --- 12.7 & 1.9 --- $\infty$\nl
Second Oldest    & 7.8 --- 9.6  & 1.1 --- 1.9 \nl
Second Youngest  & 5.7 --- 7.8  & 0.6 --- 1.1 \nl
Youngest         & 0 --- 5.7   & 0 --- 0.6 \nl
\enddata
\end{deluxetable}

\stepcounter{thetabs}
\begin{deluxetable}{cccccc}
\tablecolumns{6}
\tablenum{\tabnum}
\tablecaption{\label{momtable}
Moments of total mass and galaxy mass density distributions.}
\tablehead{Top hat radius ($h^{-1}$ Mpc)& $\sigma$ 
& $b \equiv \sigma_g/\sigma$ 
& $r \equiv \avg{\delta\delta_g}/\sigma\sigma_g$ & $\sigma_b/\sigma_g$ &
$\sigma_b/\sigma_g\sqrt{1-r^2}$ }
\startdata
1.0  & 4.77  & 2.61 & 0.886 & 0.420 & 0.905 \nl
2.0  & 2.67  & 1.94 & 0.902 & 0.367 & 0.850 \nl
5.0  & 1.14  & 1.52 & 0.923 & 0.314 & 0.817 \nl
8.0  & 0.754 & 1.41 & 0.936 & 0.295 & 0.836 \nl
16.0 & 0.401 & 1.27 & 0.941 & 0.312 & 0.920 \nl
30.0 & 0.184 & 1.24 & 0.945 & 0.303 & 0.924 \nl
\enddata
\end{deluxetable}

\stepcounter{thetabs}
\begin{deluxetable}{cccccc}
\tablecolumns{6}
\tablenum{\tabnum}
\tablecaption{\label{agetable}
Moments of total mass and galaxy mass density distributions for each 
quartile.}
\tablehead{Top hat radius ($h^{-1}$ Mpc) & Redshift Range &
$b \equiv \sigma_g/\sigma$ & $r \equiv \avg{\delta\delta_g}/\sigma\sigma_g$ 
& $\sigma_b/\sigma_g$ & $\sigma_b/\sigma_g\sqrt{1-r^2}$}
\startdata
1.0 & 1.9 --- $\infty$ & 3.58 & 0.897 & 0.372 & 0.844 \nl
& 1.1 --- 1.9 & 3.29 & 0.896 & 0.394 & 0.886 \nl
& 0.6 --- 1.1 & 2.67 & 0.745 & 0.558 & 0.836 \nl
& 0 --- 0.6 & 2.13 & 0.524 & 0.687 & 0.807 \nl
30.0 & 1.9 --- $\infty$ & 1.65 & 0.990 & 0.139 & 0.981 \nl
& 1.1 --- 1.9 & 1.56 & 0.979 & 0.188 & 0.928 \nl
& 0.6 --- 1.1 & 1.25 & 0.882 & 0.437 & 0.929 \nl
& 0 --- 0.6 & 0.834 & 0.502 & 0.817 & 0.944 \nl
\enddata
\end{deluxetable}


\stepcounter{thetabs}
\begin{deluxetable}{cc}
\tablecolumns{3}
\tablenum{\tabnum}
\tablecaption{\label{tvtable} Reduction in the standard deviation
$\sigma_{b,2}/\sigma$ for various choices of $X_i$.}
\tablehead{$X_i$ & $\sigma_{b,2}/\sigma_b$}
\startdata
Local Temperature: $T$ & 0.70 \nl
Dark Matter Velocity Dispersion: $\avg{v^2}$ & 0.68 \nl
Velocity Shear: $\Sigma/H_0^2(1+\delta)$ & 0.96 \nl
Oblateness (1 $h^{-1}$ Mpc spheres): $(\lambda_1+\lambda_2)/2\lambda_3
- 1$
& 0.97\cr
Prolateness (1 $h^{-1}$ Mpc spheres):
$2\lambda_1/(\lambda_2+\lambda_3) - 1$
& 0.97\cr
\enddata
\end{deluxetable}

%
%

\setcounter{thefigs}{1}

\clearpage

\clearpage

\clearpage
\stepcounter{thefigs}

\clearpage
\stepcounter{thefigs}

\clearpage

\clearpage
\stepcounter{thefigs}

\clearpage
\stepcounter{thefigs}

\clearpage
\stepcounter{thefigs}

\clearpage
\stepcounter{thefigs}

\clearpage
\stepcounter{thefigs}

\clearpage
\stepcounter{thefigs}

\clearpage
\stepcounter{thefigs}

\clearpage
\stepcounter{thefigs}

\clearpage
\stepcounter{thefigs}

\end{document}